\documentclass{emulateapj}
\usepackage{url}

\shorttitle{Outflows and Photoheating at High Redshift}
\shortauthors{Finlator et al.}

\begin{document}
\title{Galactic Outflows and Photoionization Heating in the Reionization Epoch}
\author{K. Finlator\altaffilmark{1,2}}
\affil{University of California Santa Barbara, Santa Barbara, CA 93106}
\altaffiltext{1}{Hubble Fellow}
\altaffiltext{2}{finlator@physics.ucsb.edu}
\author{R. Dav{\' e}}
\author{F. \"{O}zel}
\affil{University of Arizona, Department of Astronomy, Tuscon, AZ 85721}

\newcommand{\pasa}{Publications of the Astronomical Society of Australia}
\newcommand{\na}{NewA}

\newcommand{\kmsmpc}{\kms\;{\rm Mpc}^{-1}}
\newcommand{\lya}{Ly$\alpha$\ }
\newcommand{\hkpc}{h^{-1}{\rm kpc}}
\newcommand{\hmpc}{h^{-1}{\rm Mpc}}
\newcommand{\lcdm}{$\Lambda$CDM}
\newcommand{\kms}{\;{\rm km}\,{\rm s}^{-1}}

\newcommand{\mstar}{$M_{\star}$ }
\newcommand{\ud}{\mbox{\ d}}
\newcommand{\msun}{\mbox{M}_{\sun}}
\newcommand{\zsun}{\mbox{Z}_{\sun}}
\newcommand{\lgmstar}{\log(M_*/\msun)}  
\newcommand{\hinv}{h^{-1}}
\newcommand{\ebv} {$E(B-V)$}
\newcommand{\smyr} {\msun \mbox{ yr}^{-1}} 

\newcommand{\hi}{\hbox{H\,{\sc I}}}
\newcommand{\hii}{\hbox{H\,{\sc II}}}
\newcommand{\chii}{C_{\mathrm{H\,II}}}
\newcommand{\cb}{C_{\mathrm{b}}}
\newcommand{\fesc}{f_{\mathrm{esc}}}
\newcommand{\ncells}{N_{\mathrm{cells}}}
\newcommand{\tvir}{T_{\mathrm{vir}}}
\newcommand{\nhi}{n_{\mathrm{H\,I}}}
\newcommand{\nh}{n_{\mathrm{H}}}
\newcommand{\nhe}{n_{\mathrm{He}}}
\newcommand{\nhii}{n_{\mathrm{H\,II}}}
\newcommand{\nheii}{n_{\mathrm{He\,II}}}
\newcommand{\heii}{\mathrm{He\,II}}
\newcommand{\ngrid}{n_{\mathrm{grid}}}
\newcommand{\tes}{\tau_{\mathrm{es}}}
\newcommand{\zreion}{z_{\mathrm{reion}}}
\newcommand{\xmxv}{x_{\mathrm{M}}/x_{\mathrm{V}}}
\newcommand{\xhim}{x_{\mathrm{H\,I,M}}}
\newcommand{\xhiv}{x_{\mathrm{H\,I,V}}}
\newcommand{\xhiiv}{x_{\mathrm{H\,II,V}}}
\newcommand{\xhiim}{x_{\mathrm{H\,II,M}}}
\newcommand{\ngnb}{N_{\gamma}/N_b}
\newcommand{\lmfp}{\lambda_{\mathrm{MFP}}}
\newcommand{\dxr}{\Delta x_{\mathrm{R}}}
\newcommand{\dthr}{\Delta_{\mathrm{thr}}}
\newcommand{\dte}{\Delta t_{\mathrm{e}}}
\newcommand{\guvb}{\Gamma_{-12}}
\newcommand{\fcold}{f_{\mbox{\tiny cold}}}
\newcommand{\fbar}{f_{\mbox{\tiny bar}}}
\newcommand{\mh}{M_h}
\newcommand{\Cgas}{C_{\mbox{gas}}}
\newcommand{\rhodotstar}{\dot{\rho}_*}

\begin{abstract}
We carry out a new suite of cosmological radiation hydrodynamic 
simulations that explores the relative impacts on reionization-epoch 
star formation of galactic outflows and photoionization heating from a 
self-consistently grown extragalactic ultraviolet ionizing background 
(EUVB).  We compare the predictions with observational constraints 
from the cosmic microwave background, the ultraviolet continuum 
luminosity function of galaxies, and the Lyman-$\alpha$ forest.  
By itself, an EUVB suppresses the luminosity function by less than 
50\% at $z=6$ even if it is orders of magnitude stronger than observed.  
This overproduces the observed galaxy abundance by a factor of 3--5,
indicating the need for an additional feedback process.  We confirm 
that outflows readily suppress both the EUVB and the luminosity 
function into improved agreement with observations.  Population I--II 
star formation can reionize the Universe by $z=6$ even in the presence 
of strong feedback from photoheating and outflows.  The resulting EUVB 
suppresses star formation in halos with virial temperatures below 
$10^5$K but has a weaker impact in more massive halos.  Nonetheless, 
halos with virial temperatures below $10^5$K contribute up to 
$\sim50\%$ of all ionizing photons owing to the EUVB's inhomogeneity.  
Overall, star formation rate scales with halo mass $M_h$ as 
$M_h^{1.3\mbox{--}1.4}$ for halos with $M_h=10^{8.2\mbox{--}10.2}\msun$.  
This is a steeper dependence than is often assumed in reionization 
models, boosting the expected power spectrum of 21 centimeter 
fluctuations on large scales.  The luminosity function rises steeply 
to at least $M_{1600}=-13$ even in models that treat both outflows
and an EUVB, indicating that reionization was driven by faint 
galaxies ($M_{1600} \geq -15$) that have not yet been observed.  
Outflows and an EUVB 
interfere with each other's feedback effects in two ways: Outflows 
weaken the EUVB, limiting Jeans suppression of low-mass halos; this 
leads to overall de-amplification of suppression at early times ($z>8$).  
Meanwhile, they amplify each other's impact on more massive halos, 
leading to overall amplification of suppression at later times.  Our 
models cannot simultaneously explain observations of galaxies, the 
cosmic microwave background, and the intergalactic medium.  Correcting 
for dynamic range limitations and adjusting our physical treatments will 
alleviate discrepancies, but observations may still require additional 
physical scalings such as a mass-dependent ionizing escape fraction.
\end{abstract}


\section{Introduction} \label{sec:intro}
Understanding the processes that governed galaxy growth when the
Universe was less than one billion years old represents a central 
challenge for the upcoming decade.  On the largest scales, the
feedback effect of galaxies on the temperature, ionization state,
and enrichment of the intergalactic medium (IGM) will be probed 
through absorption by neutral hydrogen~\citep{fan06} and 
metals~\citep{oh02,opp09}; through its imprint on the cosmic 
microwave background~\citep{kom11}; and through redshifted emission 
by neutral hydrogen~\citep{fur06}.  On much smaller scales, 
observations of the Milky Way and its environment will decode the 
signature left by reionization in low-mass objects~\citep{bul00}.  
Between these two regimes, direct observations will constrain the 
formation of the first quasars, gamma-ray bursts, and galaxies.

With the realization that quasars could not have reionized the 
Universe by themselves~\citep{mad99,dij04b,wil10,tre11}, much 
attention is now focused on understanding the galaxies.  Using 
sensitive measurements from the Wide Field Camera 3 aboard the Hubble 
Space Telescope (HST), the Infrared Array Camera (IRAC) aboard the Spitzer 
Space Telescope, and other facilities, a number of groups have begun 
to constrain the abundance and colors of continuum-selected galaxies 
out to $z\sim10$~\citep[for example,][]{fin10,bou11b,dun11,gon11,gra11,mcl11,oes11}.  
In a complementary approach, several groups have identified tens to 
hundreds of galaxies at $z\geq6$ through their bright Lyman-$\alpha$ 
emission lines.  These catalogs have already begun to constrain the 
luminosity function, star formation, and clustering properties of faint 
galaxies~\citep{hu10,ouc10,til10,kas11}.  The James Webb Space Telescope 
(JWST) will soon push the observational frontier back to even earlier 
times~\citep{gar09}.

Interpreting these measurements as constraints on the feedback 
processes that regulated early star formation requires insight from 
theoretical models.  To date, models have demonstrated that star 
formation in low-mass halos is influenced by at least two feedback 
processes, namely galactic outflows from star-forming regions and 
photoionization heating (hereafter, photoheating) owing to an 
extragalactic ultraviolet background (EUVB).  Strong galactic outflows 
are expected in low-mass systems~\citep{dek86} but have in fact been 
observed in star-forming galaxies up to the highest masses~\citep{wei09}.  
Cosmological hydrodynamic simulations have shown that they are 
required in order to reconcile simulations with post-reionization 
observations of galaxy abundances~\citep{dav06} and scaling 
relations~\citep{dav06,fin08,dav11a,dav11b} as well as the enrichment
of the IGM~\citep{opp06,opp08}.

Photoheating impacts star formation by depleting the baryon fractions 
in low-mass halos~\citep{sha94,tho96,gne00,dij04a,oka08}.  In fact, the 
impact of an EUVB is often used to divide halos into three 
categories~\citep[for example,][]{hai09}:  Halos with virial 
temperatures in the range $300 \lesssim\tvir\lesssim10^{4}$K can use
molecular hydrogen to cool their gas to star-formation densities, but 
they are susceptible to feedback from a background in the 
H$_2$-dissociating Lyman-Werner bands.  
Halos with $10^{4}\lesssim\tvir\lesssim10^5$K can cool their gas 
through collisional excitation of neutral hydrogen, but only in the 
absence of a Lyman continuum background.  More massive halos can grow 
galaxies even in the presence of an EUVB.  For the purposes of this 
paper, we shall refer to halos with $\tvir \leq 10^4$K (virial 
velocities of $\leq 20\kms$) as ``minihalos"; halos with 
$\tvir=10^{4\mbox{--}5}$K (20--64$\kms$) as ``photosensitive", and 
more massive halos as ``photoresistant".

High-resolution radiation hydrodynamic simulations now indicate that 
a single supernova can unbind a minihalo's entire gas 
reservoir~\citep{kit05,gre07,wis08b}.  The contribution of minihalos 
to cosmological reionization is therefore probably weak, although 
their contribution to the metal enrichment of star-forming gas may be 
important~\citep{wis08a}.  Photoresistant halos are simpler 
to model because they respond weakly to an EUVB.  This means that the 
largest uncertainty in our understanding of reionization
involves the amount of star formation in the abundant but fragile 
photosensitive halos.  If the inhomogeneous EUVB does not reach a
significant fraction of photosensitive halos until late times, then 
they could have dominated the star formation density~\citep{bar00} 
and hence the ionizing photon emissivity~\citep{cho07,mun11} 
throughout much of the reionization epoch.  

The efficiency of feedback in photosensitive halos is also important 
because of its indirect impact on the more massive halos into which they 
merge.  In particular, feedback influences not only the lowest-mass 
system that is permitted to form stars, but also the way in which the 
star formation rate (SFR) scales with the halo mass ($M_h$) at higher
masses.  This hierarchical filtering is not accounted for in idealized 
models~\citep{tho96,dij04a}, but it is important because the resulting
SFR-$M_h$ scaling, when modulated by the dependence of the ionizing
escape fraction $\fesc$, determines the expected power spectrum of 
21-centimeter fluctuations at a given ionization state~\citep{mcq07}.

As such, it is a key ingredient in numerical simulations of 
reionization.  Over the past decade, a number of groups 
have used simple models for star formation in dark matter halos to 
model the reionization of representative cosmological volumes 
(\citealt{sok03,gne06,koh07,tra07,tra08,ili07,aub10}; 
see also the review in~\citealt{tra09}).  These groups have achieved 
impressive success in relating the growth of structure across a wide 
dynamic range to observational constraints on reionization from the 
cosmic microwave background (CMB) and the Lyman-$\alpha$ forest.
However, their predictions remain dependent on the assumed conversion 
from $M_h$ to SFR~\citep{mcq07}, and few of them have confronted 
post-reionization observations that constrain this scaling such as 
the IGM temperature or the galaxy LF (see, however,~\citealt{zhe10}
and~\citealt{pet10}).  

Ideally, models that use galaxies to reionize the Universe should be 
tested against observations of galaxies.  For the foreseeable future, 
the two principal constraints on star-forming galaxies will be the 
Lyman-$\alpha$ luminosity function (LF) of Lyman-$\alpha$ emitters 
(LAEs) and the ultraviolet (1350-1600 \AA) continuum LF of Lyman-break 
galaxies (LBGs).  Efforts to date have focused more on modeling the 
LAE LF because its observed evolution should be sensitive to the late 
stages of reionization~\citep{mal06,mcq07,ili08}.  
Unfortunately, this exercise depends on assumptions regarding both the
star formation efficiency within halos of different masses and the 
escape fraction of Lyman-$\alpha$ photons~\citep{zhe10}.  Estimating
the contribution of LAEs to reionization depends additionally on the 
unknown escape fraction of ionizing photons.  In short, it is easier 
to model the impact of incomplete reionization on an assumed intrinsic 
LAE LF than to constrain star formation based on the observed one.

Using the ultraviolet continuum LF (hereafter, the LF) to constrain 
the galaxy contribution to reionization has to date been prevented by 
two concerns.  First, the LF is an uncertain tracer of star formation 
owing to the unknown amount of dust extinction.  Recent work has 
demonstrated that galaxies at $z=6$ have quite blue UV continua, 
suggesting that they are relatively dust-free~\citep{bou10a,fin10,dun11}.  If 
true, then inferring SFRs of LBGs from their 1350--1600 \AA~luminosities 
is less uncertain at $z\geq6$ than at lower redshifts.  The uncertainty 
in the ionizing emissivity of galaxies then reduces to a single parameter, 
the ionizing escape fraction.  This parameter remains uncertain, but 
observations now suggest that it was larger at early times than at 
present~\citep{sia10}, and may have been as high as 50\%~\citep{rau11}.  
The second obstacle was the small available sample sizes at $z\geq6$.  
Recently, however, observations have begun to constrain the 
reionization-epoch LF~\citep{mcl11,bou11a}.  There is now a broad 
consensus as to the abundance of bright galaxies at $z=7$~\citep{mcl11}; 
constraints at $z=8$ have emerged; and star-forming galaxy candidates 
have even been identified at $z=10$~\citep{sta07,bou11a,oes11}.  These 
observations are 
currently being reinforced by even more complete samples from the 
Cosmic Assembly Near-Infrared Deep Extragalactic Legacy Survey 
(CANDELS;~\citealt{gro11},~\citealt{koe11}).  They do not yet directly constrain star 
formation in photosensitive halos, whose galaxies remain fainter than 
current detection limits.  However, the blue observed UV continua 
indicate that they constrain the star formation within photoresistant 
halos fairly directly, which may be invoked as an indirect constraint 
at fainter luminosities.

Progress in understanding the significance of photosensitive halos
requires a model for galaxy evolution that can be tested against the 
extensive observational constraints from the post-reionization Universe, 
and that can be extrapolated into the reionization epoch through the 
incorporation of a self-consistently grown inhomogeneous EUVB.  Such 
a model would treat not only the spatial inhomogeneity of 
photoheating, but also the nonlinear couplings between 
different feedback processes~\citep{pie07,paw09}.  

Cosmological hydrodynamic simulations represent a mature theoretical 
model for galaxy evolution that can serve as a starting point.  Over 
the last decade, numerous studies have shown that the 
adoption of theoretically- and empirically-motivated models for 
galactic outflows brings their predictions into reasonable agreement 
with a wide variety of observations of the IGM~\citep{opp06,opp08} 
and of galaxies~\citep{dav06,fin08,fin11,dav11a,dav11b}.  
These studies have generally assumed a spatially-uniform, optically thin 
EUVB such as that of~\citet[][hereafter HM01]{haa01}.  In order to adapt this framework 
for the reionization epoch, we recently developed a time-dependent 
continuum radiative transfer technique that is optimized for cosmological 
volumes~\citep{fin09a}.  We used this method to model the growth of 
ionized regions on snapshots extracted from existing simulations and 
verified that outflows leave enough star formation to reionize the IGM 
by $z=6$~\citep{fin09b}.  We have since integrated this method into our 
custom version of {\sc Gadget-2}, enabling us to extend our
previous work into the reionization epoch with improved realism.

In this study, we use this machinery 
to take a step toward the assembly of a complete understanding of how 
star formation may have powered reionization by modeling the relative 
roles of photoheating and outflows within the context of 
three dimensional radiation hydrodynamics simulations that successfully 
complete reionization by $z=6$.  We will consider simulations with and 
without each of these feedback processes, for a total of four kinds of 
simulations.  This will allow us to explore how strong each process is 
separately as well as how they interact when treated simultaneously.

In \S~\ref{sec:gadget2}, we discuss our custom version of {\sc Gadget-2},
our radiation transport solver and our approach to measuring
galaxy and halo properties from simulation snapshots.
In \S~\ref{sec:feedback1}, we explore how an 
EUVB and galactic outflows impact the baryon mass fractions and star 
formation rates of halos, both separately and when considered together.  
We then integrate over all halos in order to show how they impact the 
volume-averaged ionizing emissivity and star formation rate density 
(SFRD).  In \S~\ref{sec:feedback2}, we map these predictions into 
observable space by exploring the normalization and shape of the 
predicted galaxy LF.  In \S~\ref{sec:reion}, we discuss our simulated 
reionization histories and compare with inferences from the cosmic 
microwave background and the Lyman-$\alpha$ forest.  Finally, we 
summarize our results in \S~\ref{sec:summary}.  We use a standard
test case to evaluate our code in an appendix.

\section{Cosmological Radiative Hydrodynamic Simulations with {\sc Gadget-2}} \label{sec:gadget2}
In this Section, we describe our numerical methods.  Briefly, we 
run a suite of cosmological radiation hydrodynamic simulations
that use our custom version of {\sc Gadget-2} to treat gravity and
gas dynamics and are coupled with our custom radiation transport 
solver.  We identify galaxies in post-processing using {\sc skid} 
and dark matter halos using {\sc fof}.  Finally, we compute the 
galaxies' broadband photometric colors by convolving their star 
formation histories and metallicities with the~\citet{bru03} stellar 
population synthesis models.  We use a tophat filter of width 
100 \AA~centered at 1600 \AA~to define the UV continuum 
luminosity.

\subsection{Hydrodynamics, Star Formation and Outflows}\label{sec:gadget}
We model structure formation using our custom version of the parallel 
cosmological galaxy formation code {\sc Gadget-2}~\citep{spr05}.  This 
code implements a formulation of smoothed particle hydrodynamics (SPH) 
that simultaneously conserves entropy and energy and solves for the 
gravitational potential with a tree-particle-mesh algorithm.

Gas particles undergo radiative cooling using the processes and rates
in Table 1 of~\citet{kat96}.  We account for metal-line cooling using the 
collisional ionization equilibrium tables of~\citet{sut93}.  The cooling 
rate depends on the ionization state, hence cooling and ionizations must
be computed together.  We achieve this using a nested subcycling approach
that we describe below.  We initialize the IGM temperature and neutral 
hydrogen fraction to the values appropriate for each simulation's initial 
redshift as computed by {\sc recfast} (\citealt{won08}; see 
\S\ref{sec:sims}), and we assume that helium is initially 
completely neutral.

As gas particles cool, they acquire a subgrid two-phase interstellar 
medium consisting of hot gas that condenses via a thermal instability 
into cold star-forming clouds, which are in turn evaporated back into the 
hot phase by supernovae~\citep{mck77}.  They also acquire the ability to
stochastically spawn star particles via a Monte Carlo algorithm.  This 
treatment requires only one physical parameter, the star formation 
timescale, which is tuned to reproduce the~\citet{ken98} 
relation~\citep{spr03}.  The physical density threshold for star 
formation is 0.13 cm$^{-3}$.  We account for metal enrichment owing 
to supernovae of Types II and Ia as well as asymptotic giant branch 
stars; see~\citet{opp08} for details of the implementation.

We model galactic outflows (hereafter, ``outflows") using a Monte 
Carlo algorithm that applies kicks to star-forming SPH particles.  We 
tune the probability that particles are kicked and the kick velocity
such that the resulting outflow mass loading factor and wind speeds 
follow the scalings expected for momentum-driven winds~\citep{mur05}.  
In this model, the ratio of the rate at which material enters the
outflow to the SFR $\eta_{\mathrm{W}}$ varies with 
the host halo velocity dispersion $\sigma$ as 
$\eta_{\mathrm{W}} = \sigma_0/\sigma$ where the outflow amplitude 
$\sigma_0=150\kms$ is a free parameter and $\sigma$ is estimated 
during the simulation using an on-the-fly group finder.  The outflow 
velocity is proportional to $\sigma$.  This model broadly reproduces
observations of metals in the high-redshift IGM~\citep{opp08,opp10}.

\subsection{Photoionization Feedback}\label{sec:march}
\subsubsection{Discretization}
Our radiative transport (RT) solver discretizes the radiation field 
on a three-dimensional Cartesian grid and is hence Eulerian.  By 
contrast, {\sc Gadget-2} discretizes the fluid equations using 
Smoothed Particle Hydrodynamics (SPH) and is hence Lagrangian.  We 
now describe how we translate between these two discretizations.

The conversion from the SPH field to the RT grid occurs when we compute
the emissivity field owing to star-forming regions and the opacity field 
owing to partially ionized and neutral gas.  The volume-weighted mean 
emissivity $\eta$ in an RT cell is given by an integral over the cell's 
volume $V$:
\begin{eqnarray}\label{eqn:emiss1}
\eta = \frac{1}{V}\int_V \eta(\mathbf{r}) d\mathbf{r}.
\end{eqnarray}
We derive $\eta(\mathbf{r})$ from the SFRs of 
star-forming gas particles.  We account for the metallicity dependence 
of each particle's emissivity using the following fitting function:
\begin{eqnarray}\label{eqn:Q}
\log(Q) = 0.639 (-\log(Z))^{1/3} + 52.62 - 0.182
\end{eqnarray}
Here, Q is in $s^{-1} (\msun \mbox{yr}^{-1})^{-1}$, $Z$ is the metal
mass fraction, and the last term converts to a~\citet{cha03} initial
mass function (IMF).  This function reproduces the equilibrium 
emissivities of~\citet[][Table 4]{sch03} to 15\% throughout the range 
$Z \in [10^{-7},0.04]$.  We do not extrapolate this fit to 
metallicities outside this range.
\footnote{We have verified that doing so has negligible impact on 
our results, which suggests that the ionizing emissivity boost from 
hypothetical zero-metallicity stars in atomically-cooled halos has 
negligible impact on cosmological scales.}

Owing to the fact that gas particles are spatially extended and 
generally straddle cell boundaries, Equation~\ref{eqn:emiss1} becomes, 
for each cell, a sum over the SPH kernel-weighted volume means owing to 
the SPH particles that overlap that cell:
\begin{eqnarray}\label{eqn:emiss2}
\eta = \sum_i\int_V W(\mathbf{r}_i-\mathbf{r},h_i)\eta_i dV
\end{eqnarray}
Here, $\eta_i$ is the emissivity of particle $i$, $\mathbf{r_i}$
is the particle's position, $h_i$ is its smoothing length, and $W$ is the
SPH smoothing kernel.  We evaluate equation~\ref{eqn:emiss2} by fitting
a Gaussian to each SPH particle's smoothing kernel so that its contribution 
to each of its neighboring cells reduces to a sum over incomplete gamma 
functions.

Equations~\ref{eqn:emiss1}--\ref{eqn:emiss2} yield the total ionizing 
photon production rate within a cell,
but only a fraction $\fesc$ of these escape into the IGM.  We do not 
have the spatial resolution to model $\fesc$ self-consistently, hence
we treat it as a free parameter and set it to a uniform value of 50\%
for all halos.  We will show in Figure~\ref{fig:xHIJ} that this choice 
leads to reionization completing by $z\approx(7,6)$ in our fiducial
simulation volume (without, with) outflows.

In our previous calculations, we found that a much lower fiducial 
value of $\fesc=0.12$ led to reionization completing by 
$z=6$~\citep{fin09b}.  Two differences lead to the higher value that
we adopt here.  First, our post-processing simulations did not account
for subgrid gas clumping, hence they underestimated the
recombination rate.  In our current simulations, recombinations and 
cooling are computed directly on the SPH particles, automatically 
capturing gas clumping that occurs on scales smaller than the 
radiative transfer grid cells.  Second, the simulations on which we 
previously post-processed radiative transport calculations assumed 
the HM01 EUVB.  This background assumes a 
significant contribution from quasars and is harder than our 
self-consistently derived galaxies-only background.   The resulting 
density field, which was an input into our post-processing radiation 
transport calculations, possessed a higher temperature, significantly 
more Jeans smoothing, and a lower recombination rate than occurs in 
our newer, self-consistent calculations.  Tuning our simulations 
to achieve reionization by $z=6$ given our softer background 
therefore requires a higher $\fesc$.

Each RT cell's opacity $\chi$ is also given by a volume-weighted mean:
\begin{eqnarray}\label{eqn:opacity}
\chi = \frac{1}{V}\int_V \chi(\mathbf{r}) dV.
\end{eqnarray}
This equation is identical to Equation~\ref{eqn:emiss1} except that the
emissivity has been replaced with the opacity, hence we evaluate it in 
the same way.  The opacity $\chi(\mathbf{r})$ is the neutral hydrogen 
abundance multiplied by the absorption cross section 
$\sigma_{\mathrm{HI}}=6.30\times10^{-18}$cm$^2$.  We neglect the 
opacity owing to star-forming gas particles because their absorptions 
are implicitly taken into account through $\fesc$.  Adopting a smaller 
cross section in order to match the mean energy of ionizing photons 
(see below) would yield a more diffusive EUVB without qualitatively 
changing our results.

The translation from the RT grid back to the SPH field occurs when we
compute the photoionization and photoheating rates.  This involves, for
each SPH particle, summing the contributions owing to the radiation fields
from each of its neighboring cells.  For example, the photoionization rate
$\Gamma_i$ for SPH particle $i$ is given by the integral (over all space)
\begin{eqnarray}\label{eqn:gamma}
\Gamma_i = \int W(\mathbf{r}_i-\mathbf{r},h_i) \Gamma(\mathbf{r}) dV,
\end{eqnarray}
where $\Gamma(\mathbf{r})$ is the gridded photoionization rate.  We evaluate 
Equation~\ref{eqn:gamma} by fitting a Gaussian to each particle's
smoothing kernel, which again reduces the integral to a sum over
incomplete gamma functions.

The photoheating rate owing to the photoionization of
hydrogen $\epsilon_{\Gamma_{\mathrm{HI}}}$ is the
hydrogen photoionization rate $\Gamma_{\mathrm{HI}}$ times the
latent heat per photoionization $\epsilon_{\mathrm{HI}}$
\begin{eqnarray}\label{eqn:photoheat}
\epsilon_{\mathrm{HI}} = 
  \frac{\int 4 \pi \sigma_\nu J_\nu \frac{h(\nu-\nu_{\mathrm{LL}})}{h\nu} d\nu}
         {\int 4 \pi \sigma_\nu J_\nu \frac{1}{h\nu} d\nu},
	 \end{eqnarray}
where $J_\nu$ is the mean specific intensity and the integrals run from
the Lyman Limit $\nu_{\mathrm{LL}}$ to $\infty$.  $\epsilon_{\mathrm{HI}}$
depends on many factors including the intrinsic slope of the ionizing 
continuum and the impact of spectral hardening in the ISM and the 
IGM~\citep[e.g.,][]{tit07}.  If the simulation volume is large compared
to the photon mean free path, then all photons are absorbed and we may 
drop the frequency-dependent cross sections $\sigma_\nu$ from 
equation~\ref{eqn:photoheat} to obtain a volume-averaged 
$\epsilon_{\mathrm{HI}}$.  The EUVB may be approximated 
as a power law $J_\nu \propto \nu^{-\alpha}$ such that 
$\epsilon_{\mathrm{HI}}$ depends only on the slope $\alpha$.  $\alpha$ 
can vary between 5 for Population I stars and 1.8 for quasars, leading 
to values for $\epsilon_{\mathrm{HI}}$ in the range 0.25--0.64 Ryd.  
By applying population synthesis models to the young, metal-poor 
galaxies that dominate our EUVB (\S~\ref{sec:groups}), we find that 
their Lyman continua are characterized by $\alpha=$4--5 (see 
also~\citealt{bar01}).  This corresponds to 
$\epsilon_{\mathrm{HI}} = $0.25--0.32 Ryd.  We adopt 
$\epsilon_{\mathrm{HI}}=0.3$ Ryd, or 4.08 eV of thermal heating per 
ionization.  This value agrees well with the heating rate per 
ionization in the galaxies-only EUVB derived 
by HM01, but it is lower than the value of 0.6 Ryd assumed 
by~\citet{fur09} or the range 0.47--2.2 Ryd considered 
by~\citet{pet10}.  We have not varied $\epsilon_{\mathrm{HI}}$ 
because it is chosen for consistency with the simulation's predicted 
stellar populations (see \S~\ref{ssec:params} for a discussion).

\subsubsection{Solving the Radiative Transport Equation}

We solve for the transport of ionizing radiation using the moment 
method presented in~\citet{fin09a}.  This approach divides
the problem into two parts, a moment equation update and a long 
characteristics calculation that updates the Eddington tensors.  
We review this technique here and describe our parallelization 
strategy.

The zeroth and first angle moments of the radiation transport 
equation in cosmological comoving coordinates are:
\begin{eqnarray} 
\label{eqn:rt-moments-E}
\frac{\partial \mathcal{J}}{\partial t} & = &
  -\frac{1}{a} \vec{\nabla}_c \cdot \vec{\mathcal{F}} + 4\pi\eta - c \chi \mathcal{J} \label{eqn:mom0} \\
\label{eqn:rt-moments-F}
\frac{\partial \vec{\mathcal{F}}}{\partial t} & = &
  -\frac{c}{a} \vec{\nabla}_c \cdot (c\mathbf{f}\mathcal{J}) - c \chi \vec{\mathcal{F}} \label{eqn:mom1}  \\
\label{eqn:rt-moments-f}
\mathbf{f} & \equiv & \frac{\int \mathcal{N} \hat{n}\hat{n} d\Omega}{\int \mathcal{N} d\Omega}. \label{eqn:mom2}
\end{eqnarray}
Equation~\ref{eqn:mom0} relates the angle-averaged photon number density 
$\mathcal{J}$ to the divergence of the photon flux $\vec{\mathcal{F}}$, 
the angle-averaged photon emissivity $\eta$, and the opacity $\chi$, 
where $\chi$ accounts for attenuation owing to cosmological expansion, 
redshifting, and absorptions.  In our simulations, $\eta$ depends on
the local stellar population and $\chi$ is dominated by the opacity
of neutral hydrogen.  Equation~\ref{eqn:mom1} relates the evolving
photon flux to the divergence of the radiation pressure tensor (the
quantity in the parentheses) and $\chi$.  We close the moment hierarchy 
by relating the radiation pressure tensor to $\mathcal{J}$ using the
Eddington tensor $\mathbf{f}$.  We use Equation~\ref{eqn:mom2} 
to derive $\mathbf{f}$ from the angle-dependent photon number 
density $\mathcal{N}$, which we in turn obtain from a 
time-independent ray-casting procedure (see~\S3 of~\citealt{fin09a}).

We discretize equations~\ref{eqn:mom0}--\ref{eqn:mom1} fully 
implicitly in time in order to improve the code's stability.  We 
parallelize the solution to the resulting linear system of 
difference equations by decomposing the domain of radiative 
transfer grid cells over the set of compute nodes.  We then 
iteratively solve the portion of the linear system that corresponds 
to each subdomain and then swap information at the subdomain 
boundaries.  We obtain subdomain solutions from a single Gauss-Seidel 
iteration.  This ``Additive Schwarz Iteration"~\citep{sch1870} is 
straightforward to implement and lacks global synchronization points 
(because processors only need to swap information with processors 
that host neighboring subdomains).

The Eddington tensor $\mathbf{f}$ encodes the angle dependence of the
radiation field.  Errors in the assumed $\mathbf{f}$ lead to errors in 
the shapes of ionized regions (Figure 5 of~\citealt{fin09a}).  In cases 
where the timescale over which the emissivity field changes is long 
compared to the simulation's light-crossing time, obtaining $\mathbf{f}$ 
from a time-independent calculation does not introduce large errors.  
The timescale for structure to form is roughly a Hubble time, hence 
this ratio is small for simulation volumes with comoving side length 
$\sim10\hmpc$ volumes at $z\sim10$ (note that, throughout this work, 
we quote our simulation volumes' sizes in comoving units).  We compute 
$\mathbf{f}$ as follows: After each timestep, we evaluate 
whether $\mathcal{J}$ has changed within a cell by more than 5\%.  If 
so, then we update its Eddington tensor by casting rays to every 
source that is not obscured by an optical depth greater than 6.  We 
mimic periodic boundaries using a single layer of periodic replica 
volumes.  We parallelize this step by first gathering the updated 
opacity and emissivity fields onto all of the compute nodes and then 
assigning roughly equal numbers of updates to each compute node.  
Afterwards, we gather the updated Eddington tensors onto all of 
the compute nodes.

\subsubsection{Merging the Radiation and Hydrodynamic Solvers}

\begin{figure}
\epsscale{1.0}
\plotone{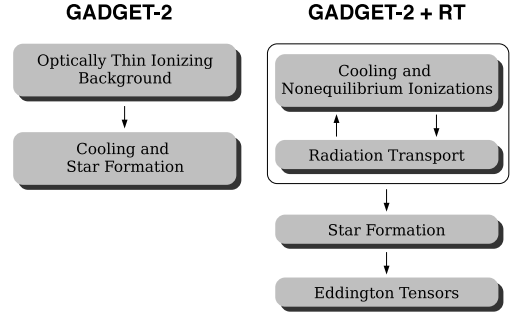}
\caption[]{Flowcharts illustrating how the conventional flow of a
single timestep in {\sc Gadget-2} (left) compares with our radiative 
hydrodynamic version (right).
}
\label{fig:flowchart}
\end{figure}

Figure~\ref{fig:flowchart} illustrates how the addition of our 
radiation transport solver modifies the flow of a single timestep in 
{\sc Gadget-2}.  In the case of a spatially-uniform, optically thin 
EUVB (left side), each timestep involves first updating the EUVB and 
then allowing the gas to cool and form stars.  In the case of 
radiation hydrodynamic simulations (right side), we first update the 
radiation, ionization, and temperature fields simultaneously.  This 
step involves three layers of nested loops.  The outermost loop 
consists of an iteration between one module that 
advances the gas ionization and thermal state and a second module 
that advances the radiation field.  Within the ionization/cooling 
module, an inner loop alternately advances the cooling equations 
explicitly and then the ionization equations implicitly using 
substeps that are limited to 0.002$u/\dot{u}$, where $u$ is the 
particle's internal energy.  The ionization solver is itself a 
Newton-Raphson iteration, which we substep separately using a constant 
temperature and ionization rate whenever any of the updated abundances 
does not converge to $10^{-4}$ in 20 iterations.  We substep the 
outermost iteration whenever the photon number density in at least one 
radiation transport cell does not converge to 10\% within 10 iterations.  
Following convergence of the outermost loop, we proceed to its next 
substep and repeat until all fields have been advanced through the 
timestep.  After updating the ionization, temperature, and radiation 
fields, we compute the number of new ``star particles" and 
``wind particles" that are spawned.  Finally, we update the Eddington 
tensor field.

Introducing additional physics into {\sc Gadget-2} requires us to slow down
the calculation in order to achieve a consistent solution.  We limit each 
particle's individual timestep based on the evolution of its electron 
abundance as follows:
\begin{eqnarray}
\Delta t \leq \left\{ 
\begin{array}{lc}
  0.1 n_e (dn_e/dt)^{-1} & n_e/n_H > 0.05 \\
  0.05 n_H (dn_e/dt)^{-1} & n_e/n_H \leq 0.05
\end{array} \right.
\end{eqnarray}
Here, $n_e$ and $n_H$ represent the electron and hydrogen number densities,
respectively.  This criterion evolves partially-ionized regions cautiously
while limiting the ability of neutral regions to slow down the calculation.
Note that our fully-implicit solution to Equations~\ref{eqn:mom0}--\ref{eqn:mom2}
obliges us to evolve the entire simulation volume's radiation field 
whenever any particle's ionization state is updated.
We also limit the global timestep to be no larger than 
$\Delta \ln(a) = 0.0035$, where $a$ is the cosmological expansion factor.

In~\citet{fin09a}, we demonstrated that our radiation transport solver 
reproduces the analytic solution for the growth of ionized regions in 
expanding media, indicating that it conserves photons and accounts accurately
for cosmological terms.  We have verified that merging our solver into 
{\sc Gadget-2} preserves this agreement.  In addition, we have tested
our merged code's ability to follow the growth of multiple HII regions 
in a realistic cosmological density field using Test 4 from the 
cosmological radiative transfer code comparison project~\citep{ili06b}.  
We show that it yields reasonable agreement with results from other 
codes in the Appendix.  

\subsection{Identifying Galaxies and Haloes}\label{sec:groups}
We isolate the galaxies that have formed within our volume at each
redshift using 
{\sc SKID}\footnote{\url{http://www-hpcc.astro.washington.edu/tools/skid.html}}.  
We infer the age and metallicity of each galaxy's stellar population 
from the age and metallicity distributions of its star particles.  We 
compute each star particle's contribution to the galaxy's integrated 
stellar continuum by interpolating to its metallicity and age within 
the~\citet{bru03} models, and we account for dust using the~\citet{cal00} 
model with a normalization that is tied to the stellar 
metallicity~\citep{fin06}.  

We identify dark matter halos following the method of~\citet{oka08}.
First, we identify overdense regions using 
{\sc FOF}\footnote{\url{http://www-hpcc.astro.washington.edu/tools/fof.html}}
with a linking length tied to the (redshift-dependent) virial overdensity.
Next, we iteratively compute the center of mass and remove the most distant
dark matter particle until only two particles remain.  We then grow a 
sphere around the midpoint between these particles until the enclosed 
density falls to the virial overdensity.  Finally, we define the halo's 
baryon mass fraction, SFR, and ionizing luminosity using the particles 
that fall within this radius.

\subsection{Simulations}\label{sec:sims}

\begin{table}

\makeatletter 
\long\def\@makecaption#1#2{%
  \vskip\abovecaptionskip 
  \sbox\@tempboxa{#1. #2}%
  \ifdim \wd\@tempboxa >\hsize
    #1. #2\par 
  \else
    \global \@minipagefalse
    \hb@xt@\hsize{\box\@tempboxa\hfil}%
  \fi
  \vskip\belowcaptionskip} 
\makeatother

\begin{tabular}{l|cccccc}
name & L$^1$ & winds & RT grid & $M_{h\mbox{,min}}^2/\msun$ & time$^3$ & $\gamma$/H$^4$\\
\hline
r6nWnRT & 6 	& no   	& --  		& $1.4\times10^8$ & 1.2 & --\\
r6nWwRT16 & 6	& no   	& $16^3$	& $1.4\times10^8$ & 2.1 & 3.9\\
r6wWnRT & 6	& yes	& --  		& $1.4\times10^8$ & 2.3 & --\\
r6wWwRT16 & 6	& yes	& $16^3$	& $1.4\times10^8$ & 22 & 5.1\\
\hline
r6nWwRT32 & 6	& no 	& $32^3$	& $1.4\times10^8$ & 4.3 & 3.6\\
r6wWwRT32 & 6	& yes	& $32^3$	& $1.4\times10^8$ & 31 & 4.9\\
r3wWwRT32 & 3	& yes	& $32^3$	& $1.8\times10^7$ & 8.8 & 5.3\\
r3nWnRT   & 3	& no	& --		& $1.8\times10^7$ & 1.5 & -- \\
\hline
r6nWHM01  & 6   & no    & --            & $1.4\times10^8$ & 1.7 & --\\
\end{tabular}
\caption{Our simulations.  All runs use $2\times256^3$ particles.
The first four simulations explore the relative impact of ouflows 
and radiative feedback.  The next four test our sensitivity to 
resolution effects.  The last run assumes the optically-thin 
EUVB of HM01.\\
$^1$ in comoving $\hmpc$\\
$^2$ virial mass of a halo with 100 dark matter and SPH particles.\\
$^3$ computation time to $z=6$ in 1000 CPU hours\\
$^4$ ionizing photons emitted per H atom at $\xhiv=0.01$}
\label{table:sims}
\end{table}

Table~\ref{table:sims} summarizes our simulation suite.  The bulk 
of our discussion will refer to a fiducial cubical volume $6\hmpc$ 
to a side, which we simulate both with/without momentum-driven 
outflows and with/without a self-consistently evolved EUVB.
We adopt a cosmology in which $\Omega_M=0.28$, $\Omega_\Lambda=0.72$,
$\Omega_b = 0.046$, $h=0.7$, $\sigma_8 = 0.82$, and the index of
the primordial power spectrum $n=0.96$.
We generate the initial conditions using an~\citet{eis99} power
spectrum at redshifts of $z=249$ and 319 for simulations subtending
cubical volumes of side length 6 and 3 $\hmpc$, respectively.  
We smooth gravitational forces using Plummer 
equivalent comoving softening lengths of 469 and 234 $h^{-1}$ 
parsecs, respectively.  The radiative transport simulations have been 
run using Eulerian grids of $16^3$ and $32^3$ radiative transport 
cells.  We also ran simulations with eight times the mass resolution 
and one eighth the volume of our fiducial $6\hmpc$ simulations in 
order to check numerical convergence.  Finally, we have re-run our 
fiducial simulation without outflows and under the assumption of 
the optically-thin HM01 EUVB for comparison with previous 
work.  

The sixth column of Table~\ref{table:sims} indicates that our RT 
solver is fairly efficient.  The ratio of the number of sources to 
the number of RT grid cells at $z=6$ is $774/4016$ in the r6wWwRT16 
and $1526/32768$ in the r6wWwRT32 simulation.  Despite this relatively 
large number of sources, self-consistent RT does not increase the total 
computation time by more than a factor of four in the absence of 
outflows (compare the r6nWnRT and r6nWwRT32 simulations).  This is 
because our ionization/RT iteration (the top bubble in the right column 
of Figure~\ref{fig:flowchart}) solves the moments of the RT equation 
for a fixed Eddington tensor.  Computing the angular dependence of the 
ionizing background (that is, updating the Eddington tensors) is 
computationally expensive, but we do this for only a fraction of RT 
cells at each timestep.

In the presence of outflows, self-consistent RT increases the
computation time by a total factor of 10--15 because outflows 
generate a significant reservoir of dense gas that lives within one
optical depth of the nearest source.  Its brief recombination
time increases the IGM's volume-averaged recombination rate
while forcing the simulation to evolve on a shorter timestep.
Consequently, outflows boost both the computation time and the 
number of ionizing photons absorbed per hydrogen atom at overlap
(seventh column in Table~\ref{table:sims}).

\begin{figure}
\epsscale{1.1}
\plotone{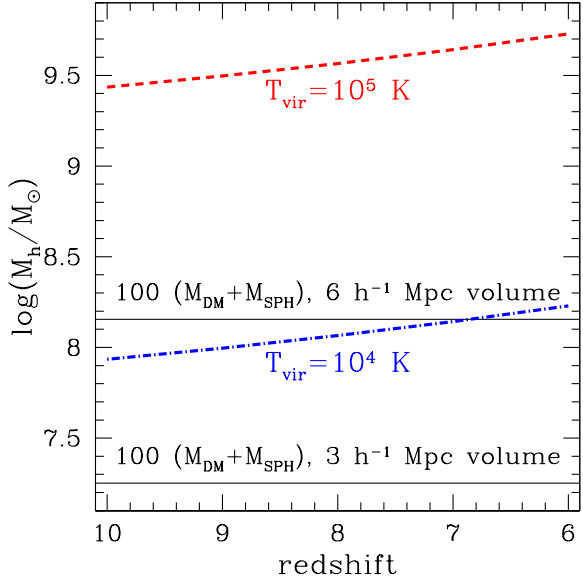}
\caption[]{The conversion from virial temperature to halo mass as
a function of redshift, computed following~\citet{bar99} with our
simulation's cosmology and a mean molecular weight of $\mu/m_p = 0.59$.  
The upper and lower solid horizontal lines indicate the masses of 
halos containing 100 dark matter and gas particles in our
6 and 3 $\hmpc$ simulations, respectively; these represent the 
threshold for halos to possess resolved masses and density 
profiles~\citep{tre10}.  Our fiducial ($6\hmpc$) volume 
resolves the HI cooling limit for all $z\leq7$.
}
\label{fig:virials}
\end{figure}

The goal of the present work is to use a three-dimensional numerical 
model to study the impact of galactic outflows and photoheating 
on star formation within photosensitive and photoresistant 
halos.  For reference, we show how this mass range varies with 
redshift in Figure~\ref{fig:virials}.  The black, solid lines 
indicate the masses of halos that contain 100 dark matter and gas 
particles in our fiducial and high-resolution simulations; halos more 
massive than these thresholds possess converged mass, virial radius, 
and density profiles~\citep{tre10}.  By design, our $6\hmpc$ 
simulations resolve halos with virial temperatures above $10^4$ K 
at $z\leq7$.  The mass 
resolution limit of our $3\hmpc$ volume includes all star formation 
in atomically-cooling halos at all redshifts, although its small 
volume limits the number of massive halos that are represented.

\section{Outflows and Photoheating I: Theoretical Insight} \label{sec:feedback1}

In this Section, we explore how outflows and photoheating
impact star formation.  We begin by analyzing the baryon mass 
fraction and SFR as a function of halo mass near the overlap 
epoch.  We will find that photoheating primarily impacts
halos below $3\times10^9\msun$, above which outflows are the dominant
feedback process.  We will show that outflows and photoheating couple
nonlinearly at all halo masses, although the sign of the effect varies 
with mass.  Next, we weight by halo abundance in order to determine how 
different halos contribute to the volume-averaged SFRD and ionizing 
emissivity.  We will show that photosensitive halos contribute 
significantly to the SFRD and the ionizing emissivity through the end 
of the reionization epoch.  Finally, we study how feedback impacts 
the evolving SFRD.

\subsection{The Halos}
\subsubsection{Baryon Fractions and Star Formation Rates} \label{ssec:halos:gen}

\begin{figure}
\epsscale{1.2}
\plotone{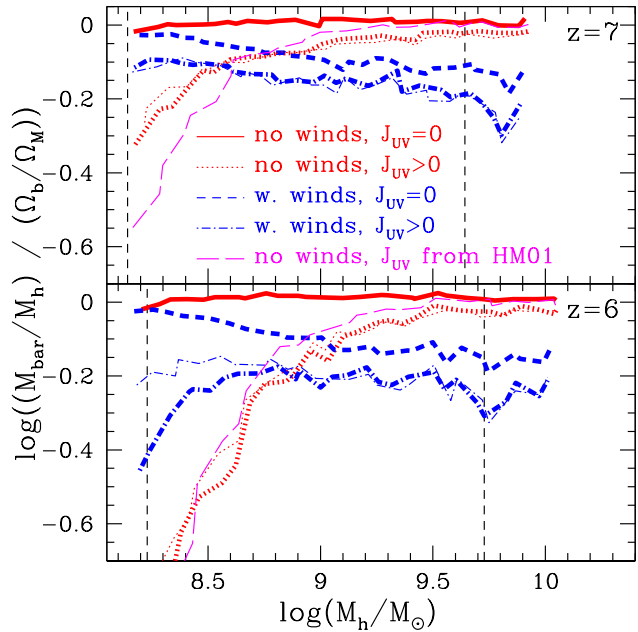}
\caption[]{The median mass fraction in baryons ($\fbar$)
normalized to $\Omega_b/\Omega_M$ as a function of halo 
mass.  Red and blue curves indicate fiducial simulations 
with/without outflows and with/without a self-consistent EUVB as 
indicated.  Thick and thin curves indicate that the EUVB is 
discretized using $16^3$ and $32^3$ RT cells, respectively.  
Magenta long-dashed curves denote simulations 
without outflows and with an optically-thin HM01 EUVB. 
Left and right vertical dashed lines mark $\tvir=10^4$K and
$10^5$K from Figure~\ref{fig:virials}.  Without feedback, halos contain 
their global baryon fraction.  An EUVB suppresses the baryon fraction 
in halos below $3\times10^9\msun$, with all baryons removed from halos 
a few times more massive than the HI cooling limit.  Including both 
feedback processes causes the baryon fraction trend to flatten at 
$0.6\Omega_b/\Omega_M$ for halos above $10^9\msun$.
}
\label{fig:fbarmhalo}
\end{figure}

In Figure~\ref{fig:fbarmhalo}, we show how the mass fraction in baryons
$\fbar$
varies with halo mass for each of our simulations at two representative 
redshifts as indicated.  The solid red curves confirm that all halos
retain their baryons in the absence of feedback.  Allowing a 
self-consistent EUVB to form (heavy red dotted) suppresses the baryon 
fraction, with strong suppression visible up to an order of magnitude 
above the HI cooling limit once the universe is reionized.  The slow
growth of $\fbar$ with mass owes to the hierarchical tendency for halos 
near the HI cooling limit to accrete much of their mass in the 
form of halos that are below the limit.  Simulations that treat outflows
but not an EUVB (heavy blue short-dashed) show that outflows suppress 
baryon fractions only in halos where star formation is efficient, hence
they are negligible at the HI cooling limit.  This is an important 
difference between outflows and an EUVB: photoheating primarily impacts 
systems near the HI cooling limit while outflows preferentially impact 
more massive systems owing to hierarchical merging.  The separate 
effects of outflows and photoheating match at roughly $10^9\msun$, 
above which photoheating weakens while outflows remain efficient by 
design.  Finally, treating both outflows and an EUVB suppresses the 
baryon fraction at all sampled masses.  Curiously, adding galactic 
outflows to a simulation that already grows an EUVB (that is, going 
from the dotted red to the dot-dashed blue curves) \emph{boosts} the 
baryon fraction below $10^9\msun$.  We will return to this 
``suppression of suppression" below.

In order to compare our predictions to previous work, we show with a 
magenta long-dashed curve the baryon fraction that results without winds 
and with the optically thin HM01 EUVB.  Comparing
this with the dotted red curves shows that a spatially-resolved EUVB 
yields higher baryon fractions at low masses at $z=7$.  Low mass halos
have only just been exposed to the EUVB in this simulation because it
completes overlap around $z=7$ (Figure~\ref{fig:xHIJ}), hence their 
baryon reservoirs have not yet responded.  This illustrates the 
sensitivity of the predicted baryon fractions to inhomogeneous 
reionization scenarios as compared to models that assume a homogenous 
EUVB.  By $z=6$, the resulting baryon fractions are within 10\% of each 
other even though the amplitude of the HM01 EUVB is an order 
of magnitude weaker (Figure~\ref{fig:xHIJ}).  This confirms that 
baryon mass fractions depend weakly on the EUVB amplitude, as opposed 
to its bias and hardness.  This can be understood as follows: A halo's 
baryon mass fraction depends mostly on its virial temperature relative 
to the temperature of the surrounding IGM.  The IGM temperature depends 
mostly on the EUVB hardness at the moment it was reionized because 
it decouples from the EUVB afterwards.  Therefore, boosting the 
EUVB amplitude in an ionized region does not impact the local 
baryon fractions because it does not impact the IGM temperature 
(see also~\citealt{mes08}).

\begin{figure}
\epsscale{1.2}
\plotone{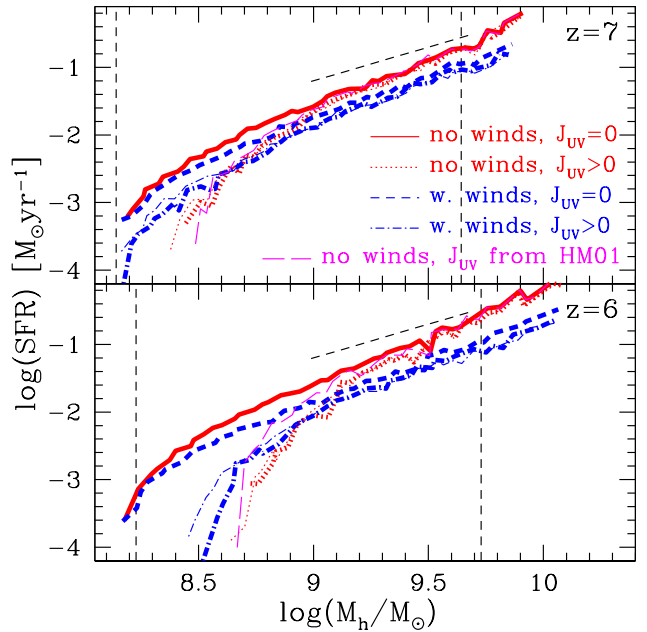}
\caption[]{SFR as a function of halo mass.  Simulations and line 
styles are as in Figure~\ref{fig:fbarmhalo}.  Star formation is 
suppressed in halos that approach the HI cooling limit (left vertical 
dashed line) even though these halos accrete their cosmic 
complement of baryons.   At higher masses, the relation between 
SFR and halo mass approaches $\mbox{SFR} \propto M_h$.  Overall, 
halos that are baryon-depleted also show suppressed star formation 
rates as expected.
}
\label{fig:sfrmhalo}
\end{figure}

In Figure~\ref{fig:sfrmhalo}, we show how SFR varies 
with halo mass in the same simulations.  In all cases, star formation 
is negligible in halos below the HI cooling limit because minihalos
cannot cool by collisionally exciting neutral hydrogen.  Above this 
cutoff, SFRs rise superlinearly over roughly a decade 
in mass before asymptotically approaching a high-mass trend that is 
generically steeper than $\mbox{SFR} \propto M_h$ (black long-dashed line 
segment with arbitrary normalization).  An EUVB raises the cutoff 
mass roughly a factor of 3 above the HI cooling limit by suppressing 
the baryon fractions in photosensitive halos (Figure~\ref{fig:fbarmhalo}).  
Star formation in halos more massive than $3\times10^9\msun$ is weakly 
affected by the EUVB.  Assuming an optically-thin HM01 EUVB 
and neglecting outflows results in SFRs that are $\sim10$\% higher than 
predictions from our self-consistent EUVB, reflecting slightly higher 
baryon fractions in the presence of a weaker EUVB (Figure~\ref{fig:xHIJ}).  
Overall, relative SFRs given different feedback models follow the 
trends expected from the baryon fractions.

\begin{table}

\makeatletter 
\long\def\@makecaption#1#2{%
  \vskip\abovecaptionskip 
  \sbox\@tempboxa{#1. #2}%
  \ifdim \wd\@tempboxa >\hsize
    #1. #2\par 
  \else
    \global \@minipagefalse
    \hb@xt@\hsize{\box\@tempboxa\hfil}%
  \fi
  \vskip\belowcaptionskip} 
\makeatother

\centering
\begin{tabular*}{0.4\textwidth}{@{\extracolsep{\fill}}l|cccc}
$z$ &  $a^1$  &  $b$   &  $M_c^2$  &  $\alpha$ \\
\hline
   & \multicolumn{3}{c}{r6wWnRT (winds, no EUVB)} & \\
\hline
10 & 0.56    & 1.29 & 8.08 & 3.3 \\
9  & 0.50$^3$& 1.29 & 8.15 & 3.6 \\
8  & 0.50    & 1.35 & 8.18 & 5.1 \\
7  & 0.36    & 1.29 & 8.27 & 4.0 \\
6  & 0.29    & 1.30 & 8.30 & 5.1 \\
\hline
 & \multicolumn{3}{c}{r6wWwRT16 (winds + EUVB)} & \\
\hline
10 & 0.49 & 1.27 &  8.10 &  4.6 \\
9  & 0.47 & 1.30 &  8.18 &  4.0 \\ 
8  & 0.42 & 1.36 &  8.22 &  5.1 \\
7  & 0.31 & 1.33 &  8.36 &  3.0 \\
6  & 0.25 & 1.40 &  8.61 &  3.4 \\
\end{tabular*}
\caption{Fits using Equation~\ref{eqn:sfrmh} to the simulated 
relationship between SFR and $M_h$ in models with outflows, 
and with and without an EUVB.\\
$^1$ in $\smyr$\\
$^2$ in $\log_{10}(M_c/\msun)$\\
$^3$ Owing to outliers, we were forced to set $a=0.50\smyr$ 
by hand in this model.  All other parameters were allowed
to vary freely.}
\label{tab:sfrmh}
\end{table}

We have fit the simulated trend of SFR versus $M_h$ in the wind models
both with and without an EUVB using a fitting function that consists 
of a power-law at high masses and a turnover at low masses (the
turnover uses the functional form of~\citealt{gne00}):
\begin{equation}\label{eqn:sfrmh}
\dot{M}_* = 
	\frac{
		a \left(\frac{M_h}{10^{10}\msun}\right)^b
	}{
		\left[1 + (2^{\alpha/3}-1)\left(\frac{M_h}{M_c}\right)^{-\alpha}\right]^{3/\alpha}
	}
\end{equation}
Table~\ref{tab:sfrmh} gives the parameters of the resulting fits.  In 
obtaining these parameters, we allow the normalization $a$, high-mass slope 
$b$, turnover mass $M_c$, and turnover slope $\alpha$ to vary independently 
at each redshift except as indicated in the caption.  The normalization 
$a$ is 10--20\% lower in the presence of an EUVB, and it decreases by 
40\% from z=$10\rightarrow6$.  The high-mass slope $b$ is always 
steeper than the linear relation $\mbox{SFR} \propto M_h$.  We have verified 
that our higher-resolution winds+EUVB model (r3wWwRT32; see 
Table~\ref{table:sims}) also yields $b=1.3$--1.4, hence this 
scaling is robust to mass resolution effects.  This has implications 
for the typical size of ionized regions and the expected power 
spectrum of 21 centimeter fluctuations.  For example, Figure 17 
of~\citet{mcq07} indicates that changing $b$ from 1 to 5/3 increases 
the expected power by $\sim10\times$ at 0.1 h Mpc$^{-1}$ scales if 
the neutral fraction is 20\%.  Our simulations are consistent with the 
steeper end of this range, indicating more power at large scales.  
The cutoff mass $M_c$ tracks the HI cooling mass in the absence of 
an EUVB, but photoheating boosts it by an additional 
factor of 2 towards $z=6$.  The turnover slope $\alpha$ is not 
well-constrained because the scatter grows large to low masses, but it
lies within the range 3--5.  These fits may be compared with other 
models for galaxy growth during the reionization epoch.

Resolution limitations impact our predictions in two ways.  First,
the limited spatial resolution of our radiation transport solver can 
dilute the EUVB as long as ionized regions are not large compared to the 
grid cells, oversuppressing unbiased halos whose environments would remain 
neutral for longer at higher resolution.  We show this in 
Figure~\ref{fig:fbarmhalo} by using heavy and light curves to compare 
results from our fiducial $6\hmpc$ volume when the RT grid cells span 375 
and 187.5 $\hkpc$, respectively.  At both redshifts, the no-wind simulation 
is insensitive to spatial resolution because it completes reionization 
at $z=7$.  By contrast, the simulation that includes outflows and an EUVB 
is just completing reionization at $z=6$.  In this case, halos below 
$10^9\msun$ retain more baryons at higher spatial spatial resolution 
because their (relatively unbiased) environments reionize later.  The
impact on the predicted SFRs is as expected in Figure~\ref{fig:sfrmhalo}.

The second resolution limitation involves the limited mass and spatial
resolution of our hydrodynamics solver.  Our fiducial simulation misses
some star formation at redshifts $z>7$ because the HI cooling mass
falls below its 100-particle resolution limit, which delays
reionization.  For similar reasons, gas clumping is not resolved
at densities above the limit given by the SPH smoothing length, 
which delays star formation in all halos.  We evaluate these limitations 
using the r3wWwRT32 simulation, which trades decreased volume for 
increased mass resolution.  Note that this comparison gives an 
incomplete accounting of resolution limitations because shrinking the 
cosmological volume delays the growth of structure owing to the lack 
of long-wavelength density fluctuations~\citep{bar04} while
increasing the mass resolution accelerates the growth of stellar 
mass owing to the predominance of small structures at early 
times~\citep{spr03}; these effects can partially cancel.  A more 
complete accounting will require increased dynamic range, which is
a goal for future work.

\begin{figure}
\epsscale{1.1}
\plotone{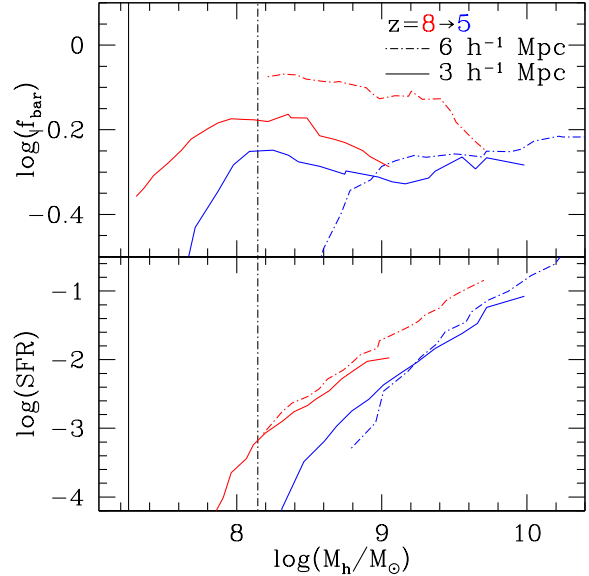}
\caption[]{\emph{(top)} The impact of mass resolution on $\fbar$ at 
$z=8$ (upper red) and $z=5$ (lower blue).  Dot-dashed and solid curves 
refer to our r6wWwRT16 and r3wWwRT32 simulations, respectively (see 
Table~\ref{table:sims}), while the vertical line segments indicate 
the 100-particle mass resolution limits.  \emph{(bottom)} The impact 
of resolution limitations on SFR.  The units of the y-axes are as in
Figures~\ref{fig:fbarmhalo}--\ref{fig:sfrmhalo}.  We have smoothed
all curves with a tophat of width 0.2--0.3 magnitudes.  Increasing mass 
resolution by a factor of 8 suppresses $\fbar$ and SFR by a factor of 
less than $\lesssim50$\%.
}
\label{fig:fbarsfrmhalor3r6}
\end{figure}

In the top panel of Figure~\ref{fig:fbarsfrmhalor3r6}, we show how $\fbar$ 
evolves during $z=8\rightarrow5$ in our fiducial (r6wWwRT16, dot-dashed) and 
high-resolution (r3wWwRT32, solid) simulations.  At $z=8$ (upper red curves), 
increasing the mass resolution by a factor of eight reduces the baryon 
fractions by 0.1 dex.  That this occurs even though the IGM is still 
essentially neutral in both
simulations (Figure~\ref{fig:xHIJ}) indicates that the dominant effect is 
the ability of simulations with higher mass resolution to resolve star 
formation and drive outflows sooner, drawing down their baryon reservoirs.  
By $z=5$, when both simulations have completed reionization and the fiducial
simulation has been forming stars for longer, the predicted baryon fractions 
at $10^9\msun$ agree well.  Below $10^9\msun$, the high-resolution simulation 
predicts higher baryon fraction because its small volume leads to a delayed 
overlap epoch~\citep{bar04}; in essence, halos at the HI cooling mass have 
only just seen an EUVB in the high-resolution simulation whereas they have
seen one for many dynamical times in the fiducial volume.

In the bottom panel of Figure~\ref{fig:fbarsfrmhalor3r6}, we show that 
the offsets in $\fbar$ translate into offset SFRs as expected: At $z=8$, 
the high-resolution simulation has a slightly suppressed SFR while at 
$z=5$ its extrapolation to high masses roughly agrees with the trend 
in the fiducial volume.  Halos less massive than $10^9\msun$ have 
significantly higher SFRs in the high-resolution simulation because 
their baryon reservoirs have not yet responded fully to reionization.  
In both cases, the SFR vanishes near the HI cooling mass.

Overall, resolution limitations in our fiducial simulation volume lead to 
errors of order 0.1 dex in $\fbar$ and SFR, particularly in halos 
less massive than $10^9\msun$.  The important point is that they do not 
change the qualitative evolutionary trends.  We conclude that our fiducial 
simulation volume contains enough dynamic range to capture the relative 
impacts of different feedback effects even though absolute predictions will 
have to await the arrival of simulations that sample a larger dynamic 
range.

\subsubsection{Nonlinear Coupling of Feedback Mechanisms by Halo Mass} \label{ssec:halos:nonlin}

\begin{figure}
\epsscale{1.1}
\plotone{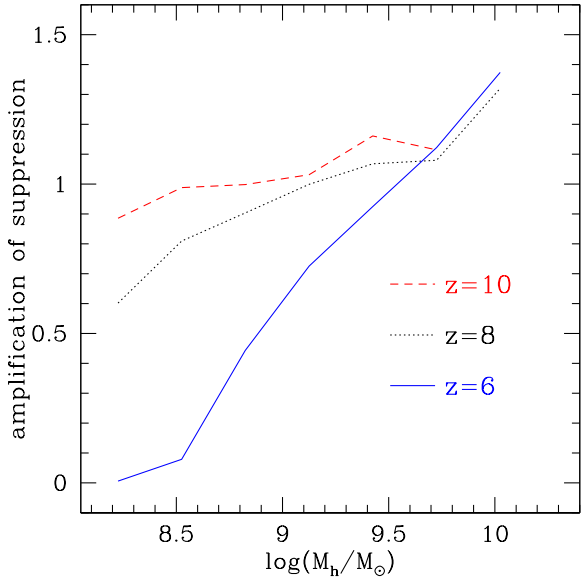}
\caption[]{Nonlinear amplification of SFR suppression as a function of 
halo mass in our fiducial volume at three different redshifts as indicated.  
Suppression of the EUVB by outflows weakens suppression of star formation 
in photosensitive halos.  Coupling between outflowing material and an EUVB 
boosts the suppression of star formation in photoresistant halos.
}
\label{fig:sfrmhaloamp}
\end{figure}

Comparing the blue dot-dashed and red dotted curves in 
Figures~\ref{fig:fbarmhalo}--\ref{fig:sfrmhalo} reveals that accounting 
for both outflows and photoheating leads to higher baryon fractions and 
SFRs below $\sim10^9\msun$ than in EUVB-only models.  This suppression 
of suppression has two causes.  First, outflows suppress the EUVB 
(Figure~\ref{fig:xHIJ}), leading to a cooler IGM with a lower Jeans 
length that permits lower-mass halos to retain their baryons.  Second, 
the simulation is 
just completing reionization at $z=6$, hence many of the lowest-mass 
halos have only recently been exposed to an EUVB and have already 
cooled their baryon reservoirs to high densities; they will continue
to form stars for many dynamical times~\citep{dij04a}.  Both effects 
weaken the impact of an EUVB on photosensitive halos at the overlap 
epoch.  They are examples of coupling between feedback processes, and 
they illustrate the role of radiation hydrodynamic simulations in the 
study of reionization.

We explore feedback coupling further in Figure~\ref{fig:sfrmhaloamp}.  
Here we have computed the ``amplification of suppression" in two steps 
following~\citet{paw09}: First, we divide the mean trend between SFR 
and halo mass from runs that include feedback by the trend from the 
no-feedback run to obtain individual ``suppression factors".  Next, 
we divide the suppression factor from the winds+EUVB model into the 
product of the suppression factors from the EUVB-only and winds-only 
models.  This yields the factor by which feedback coupling amplifies 
the total suppression of star formation over what would be expected 
if outflows and an EUVB interacted linearly.  
Figure~\ref{fig:sfrmhaloamp} indicates that the sign of feedback 
amplification depends on halo mass: The tendency of outflows to 
suppress the EUVB weakens feedback on photosensitive halos, while 
coupling between outflows and the EUVB amplifies feedback on 
photoresistant halos.  We will return to feedback coupling in
\S~\ref{ssec:rhodotstar}.  We have verified that the amplification 
factors in Figure~\ref{fig:sfrmhaloamp} are essentially unchanged 
if we double the spatial resolution of the radiation transport 
solver.

\subsection{Averaging over Cosmological Scales} \label{ssec:rhodotstar}

\subsubsection{The Differential Ionizing Emissivity} \label{sssec:rhodotstarm}

\begin{figure}
\epsscale{1.2}
\plotone{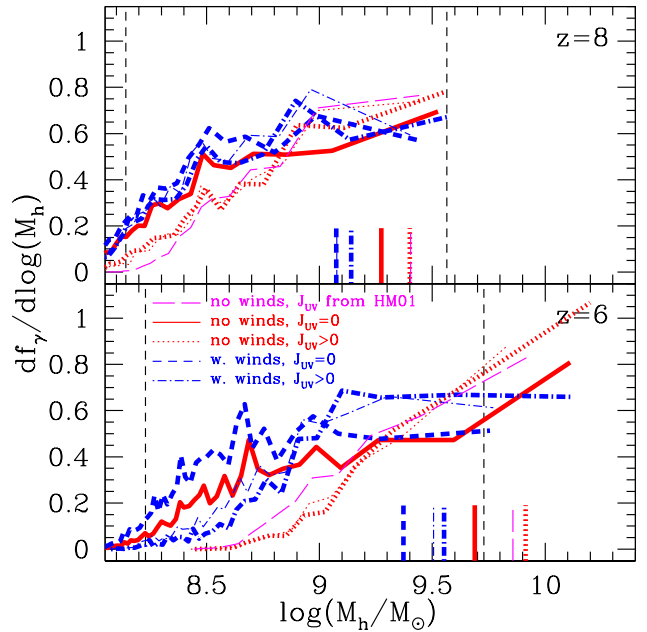}
\caption[]{The differential fraction of ionizing emissivity as a 
function of halo mass in each of our models at $z=8$ \emph{(top)} 
and $z=6$ \emph{(bottom)}.  All curves include the effect of low
metallicity in low-mass halos, and have been normalized to unit 
area.  Line colors and styles are indicated in the legend.  Heavy and 
light curves distinguish simulations with $16^3$ and $32^3$ radiation 
transport cells, respectively.  Black vertical dashed lines indicate 
virial temperatures of $10^4$K (left) and $10^5$K (right).  Vertical 
segments indicate the mass below which 50\% of ionizing emissivity 
occurs.  Photosensitive halos contribute significantly in all models, 
but particularly in the presence of outflows.
}
\label{fig:fsfr}
\end{figure}

Figures~\ref{fig:fbarmhalo}--\ref{fig:sfrmhalo} confirm that 
photoheating suppresses galaxy growth in photosensitive halos.  
However, they can still contribute significantly to 
the volume-averaged SFRD and ionizing emissivity owing to their
abundance.  We show in Figure~\ref{fig:fsfr} the fractional 
contribution by halos of different masses to the total ionizing 
emissivity at two representative redshifts.  This can be 
thought of as the integral of the product of the curves in 
Figure~\ref{fig:sfrmhalo} with the dark matter halo mass function, 
but in practice we simply sum directly over the simulated halos.  
We weight each halo's SFR by the (self-consistently predicted) 
metal mass fraction of its star-forming gas through 
Equation~\ref{eqn:Q}.

In the absence of outflows (red dotted), an EUVB systematically 
suppresses the total fraction of ionizing photons contributed by 
lower-mass halos, raising the mass below which 50\% of ionizing
emissivity occurs (vertical segments) from $10^{9.7}$ to 
$10^{9.9}\msun$.  The resulting no-outflow EUVB is orders of magnitude 
stronger than the optically-thin HM01 EUVB (magenta dashed;
see also the bottom panel of Figure~\ref{fig:xHIJ}), but it suppresses
star formation in low-mass halos only marginally more effectively (as 
previously seen in \S~\ref{ssec:halos:gen}.)
Outflows without an EUVB boost the relevance of low-mass halos because 
hierarchical merging systematically depletes the baryon reservoirs of 
more massive halos more effectively (Figure~\ref{fig:fbarmhalo}).  In 
simulations that include both forms of feedback, low-mass halos are 
more suppressed than in winds-only models owing to photoheating, but 
less suppressed than in EUVB-only models because the EUVB is weaker.

We use vertical dashed lines to bound the range of halo masses
corresponding to $\tvir=10^{4\mbox{--}5}$K, or photosensitive halos.  
Comparing the photosensitive 
mass range to the vertical segments reveals that photosensitive 
halos contribute significantly in all models at all epochs, and 
they generate more than half of all ionizing photons for all 
$z\geq6$ in models that include outflows.  This fraction is 
slightly overestimated because our fiducial volume does not sample 
the halo mass function above $2\times10^{10}\msun$.  We can
estimate the contribution of more massive halos by weighting 
the Sheth-Tormen mass 
function\footnote{kindly calculated for us in our assumed 
cosmology by J.\ Mu{\~n}oz} 
by the fits in Table~\ref{tab:sfrmh} and integrating.  Correcting
for the limitations of our finite volume in this way, we find that 
photosensitive halos contribute (49,31)\% of all star formation at 
$z=(7,6)$ in our fiducial winds+EUVB model.  This simple estimate 
neglects the weak dependence of metallicity on halo mass as well as 
slight inaccuracies at the massive end of the Sheth-Tormen mass 
function~\citep{luk07}.  Nevertheless, it confirms that, for 
realistic inhomogeneous reionization scenarios, the abundance of 
photosensitive halos more than compensates for their susceptibility 
to outflows and photoheating.

We have created a galaxy analog to Figure~\ref{fig:fsfr} in which 
the x-axis is the absolute magnitude $M_{1600}$.  Summing over
all simulated galaxies, we find that, in our 
favored winds+EUVB model, half of all ionizing photons originate in 
galaxies fainter than $M_{1600}\approx-15$.  This luminosity is 
roughly ten times fainter than the deepest current constraints at 
$z=6$~\citep{bou11b}.  It lies close to the resolution limit below 
which star formation is artificially suppressed (Figure~\ref{fig:lfUVr3r6}),
hence it is probably biased bright.  These considerations reinforce 
the need for fainter observations to constrain the dominant 
population of ionizing sources at $z>6$.

\subsubsection{The Volume-Averaged Star Formation Rate Density} \label{sssec:rhodotstar}

\begin{figure}
\epsscale{1.05}
\plotone{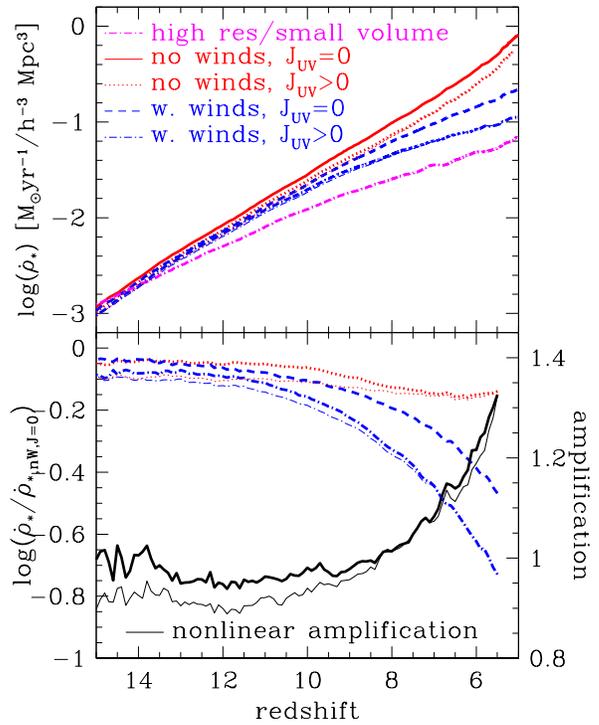}
\caption[]{\emph{(top)} The volume-averaged total star formation rate 
density (SFRD) as a function of redshift.  \emph{(bottom)} The ratio 
of the curves in the top panel to the curve from the simulation that 
includes neither outflows nor an EUVB.  Line colors and styles are 
indicated in the legend.  Heavy and light curves distinguish simulations
with $16^3$ and $32^3$ radiation transport cells, respectively.  
Both outflows and an EUVB suppress the SFRD.  The solid black curve 
illustrates feedback amplification.  The combined suppression is less 
than the sum of their separate effects at early times and greater at 
late times.  
}
\label{fig:madau}
\end{figure}

In Figure~\ref{fig:madau}, we show how the volume-averaged star 
formation rate density (SFRD) varies with redshift in each of our 
simulations.  The solid red curve in the top panel illustrates how 
the predicted SFRD grows in simulations without any feedback.  
Comparing the dotted red and dashed blue curves reveals that 
outflows and photoheating suppress the SFRD by comparable
factors.  The dot-dashed blue curve lies significantly below the 
other curves, indicating that both processes are important in 
radiation hydrodynamic simulations.  Note that the predicted SFRD 
suppression is expected to be robust to mass resolution at $z\leq7$ 
because our simulations resolve the HI cooling limit 
(Figure~\ref{fig:virials}).  At earlier times, they underestimate 
the impact of photoheating because they only partially account for 
star formation in halos below $1.4\times10^8\msun$.  Our 
high-resolution/small-volume simulation has a lower SFRD than the 
$6\hmpc$ volume following $z=13$ because it lacks sources brighter 
than $M_{1600}=-15$ (Figure~\ref{fig:lfUVr3r6}). 

We quantify the amount by which feedback suppresses each simulation's 
SFRD in the bottom panel.  By itself, the EUVB suppresses the SFRD 
by a factor that grows to $\approx30\%$ by $z=6$.  Its impact flattens 
below $z=7$ because photoresistant halos come to dominate the SFRD.  
By contrast, the impact of outflows grows with time because the 
timescale on which galaxies of any mass replace baryons that they 
eject grows as the cosmic density declines.  Put differently, winds 
cause the SFR to be dominated by the gas inflow rate at earlier 
redshifts than in models that ignore outflows.  By $z=6$, outflows 
reduce the SFRD by $\approx60\%$.  Finally, treating an EUVB and 
outflows simultaneously suppresses the SFRD by an amount that grows 
to $\approx75\%$ by $z=6$.

In both panels of Figure~\ref{fig:madau}, we use heavy and light
curves to compare the results from using coarse and fine grids in
our radiation transport solver.  The resulting curves appear 
coincident in the top panel, but the bottom panel clearly shows 
that lower spatial resolution results in decreased suppression of 
the SFRD at early times owing to the tendency for coarse grids to 
dilute the EUVB.  The effect is only noticeable while the ionized
regions are not large compared to the grid cells, however, and
by $z=8$ it is much smaller than the systematic differences between
different models.

Our simulations predict somewhat more suppression by
$z=6$ than~\citet{pet10}, who find 0--10\% depending on their 
parameter choice (their Figures 10--11).  The difference likely owes 
predominantly to the fact that the mass resolution of our simulations
is $37\times$ as high as in their fiducial volume, a choice that 
reflects our emphasis on feedback in low-mass halos rather than the 
IGM.  Our high resolution allows us to model the impact of an EUVB 
on the abundant low-mass halos that fall below their resolution and
that dominate star formation at early times (at the expense of
subtending a smaller cosmological volume).  An additional contribution
to the difference may owe to the fact that~\citet{pet10} resolve 
self-shielding filaments more effectively than in our simulations, 
which would weaken the suppression of low-mass halos in their
calculations.

Figure~\ref{fig:madau} suggests that the volume-averaged suppression 
factor from treating outflows and an EUVB simultaneously differs 
from the sum of their separate impacts.  This could occur because
an EUVB injects entropy into inflowing gas, increasing its cross 
section to entrainment by outflows~\citep{kit05}.  Alternately, 
it could occur because outflows ``puff up" overdense gas, increasing 
its cooling time and rendering it more susceptible to photoheating.  
It is expected given Figure~\ref{fig:sfrmhaloamp}, but its overall
sign and magnitude depend on the predicted SFR-$M_h$ scaling.

In order to compute the volume-averaged impact of feedback coupling, 
we multiply the EUVB-only (red dotted) and winds-only (blue dashed) 
suppression factors to obtain the predicted total suppression factor 
assuming no feedback amplification (not shown).  We then divide this 
into the true total suppression factor (blue dot-dashed) to find the 
feedback amplification factor (solid black, right y-axis).  
Overall amplification is below unity at early times, indicating that 
the tendency for outflows to suppress the EUVB, thereby boosting 
star formation in low-mass halos, wins over any more direct coupling 
between outflows and the EUVB.  This was anticipated in 
Figures~\ref{fig:fbarmhalo}--\ref{fig:sfrmhalo}, where we showed 
that, if the EUVB is grown self-consistently, then photosensitive halos 
possess higher baryon fractions and SFRs with outflows than without.  
The effect is stronger if the radiation transport solver uses higher 
spatial resolution because this further confines the EUVB to the most 
overdense regions, weakening its impact in voids.  Once reionization
is well under way, amplification grows because star formation is 
increasingly dominated by photoresistant halos.  Amplification 
exceeds 20\% by $z=6$ and continues growing into the post-reionization 
epoch. 

Figures~\ref{fig:sfrmhaloamp} and~\ref{fig:madau} may be compared
with the results of~\citet{paw09}, who studied feedback coupling 
using a suite of simulations with/without outflows and with/without 
an optically-thin EUVB.  They found that photoheating and outflows 
positively amplify each other's effects at all redshifts.  Our study 
builds on theirs in a number of ways.  First, we model an 
inhomogeneous EUVB, which qualitatively accounts for the possibility 
that the overdense regions that feed gas into halos could self-shield 
against the EUVB until relatively late times~\citep{fin09b}.  In detail, 
our simulations do not yet completely resolve self-shielding because our 
RT cells are wider than the length scales of the Lyman Limit systems 
that host photosensitive halos ($\sim10$ physical kpc;~\citealt{sch01}).
This explains why the baryon fractions in photosensitive halos are not 
converged in Figure~\ref{fig:fbarmhalo}.  It will be important 
in future work to model self-shielding within
resolution-convergent calculations.  Second, 
we grow the EUVB self-consistently, which accounts for the possibility 
that outflows suppress the EUVB by suppressing star formation 
(Figure~\ref{fig:xHIJ}).  Finally, we account for metal-line cooling.  
Metal ions enhance the cooling rate of enriched gas, weakening its 
response to an EUVB.  For the same reason, they boost the SFRD and 
hence the amplitude of a self-consistent EUVB.  Although we do not 
isolate the impact of metal-line cooling individually, we confirm 
that it qualitatively preserves feedback coupling.

Quantitatively, our Figure~\ref{fig:madau} suggests an overall 
positive amplification of 20\% by $z=6$, as compared to 60\% in
Figure 1 of~\citet{paw09}.  We attribute the difference primarily 
to the tendency for outflows to suppress the EUVB, leading to 
``de-amplification" of suppression in photosensitive halos 
(Figure~\ref{fig:sfrmhaloamp}).  Nonetheless, we qualitatively 
confirm their primary result that hydrodynamic and radiative 
feedback effects couple nonlinearly.  Our predicted amplification 
would probably converge with theirs at lower redshifts although we 
have not evolved our simulations past  $z=5$.

\section{Outflows and Photoheating II: Observational Implications} \label{sec:feedback2}

In this section, we study how outflows and photoheating impact 
the UV continuum LF of galaxies during the 
reionization epoch.  We will show that simulations without outflows
overproduce the observed LF while simulations that include outflows
are in reasonable agreement.  In both cases, an EUVB suppresses the LF 
normalization by less than 30\%.  Meanwhile, the LF is not expected 
to flatten at luminosities brighter than $M_{1600} = -13$, roughly
a factor of 100 fainter than current observational limits at $z=6$.

\subsection{Normalization of the Luminosity Function} \label{ssec:lfnorm}

\begin{figure}
\epsscale{1.1}
\plotone{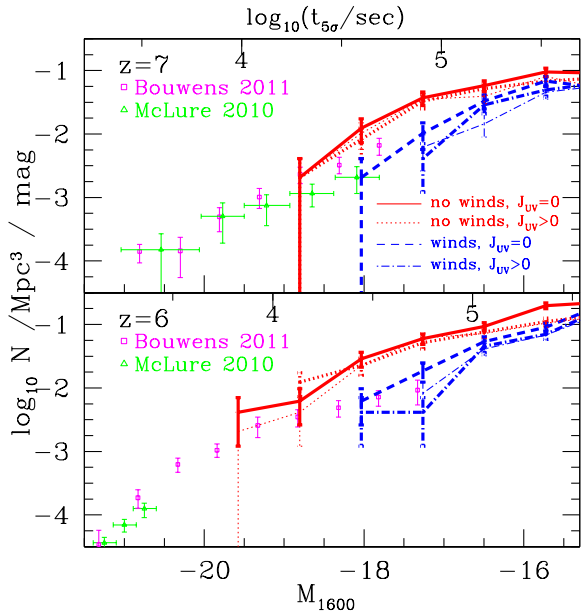}
\caption[]{The simulated rest-frame ultraviolet continuum LF
including dust for simulations with/without outflows and 
with/without an EUVB at two different redshifts (see the legend for
the line styles).  We exclude galaxies with fewer than 64 star 
particles.  Simulated errors are Poisson.  The top axis indicates the 
integration time for a $5\sigma$ detection in the F200W band of JWST 
in units $\log_{10}(t/\mathrm{sec})$.  Data are from~\citet{bou11b} 
and~\citet{mcl10}.  Both outflows and an EUVB suppress the 
normalization of the LF into improved agreement with observations at 
$z\leq7$.  Even in the presence of strong feedback, the LF continues 
to rise steeply below current observational limits.
}
\label{fig:lfUV}
\end{figure}

In Figure~\ref{fig:lfUV}, we show how outflows and photoheating
impact the predicted LF at two representative redshifts.
Intuitively, models that predict more strongly suppressed baryon 
fractions (see Figure~\ref{fig:fbarmhalo}) also predict more strongly 
suppressed LFs.  The no-feedback model predicts the highest galaxy 
abundance within the errors.  Next is the EUVB-only model, whose LF 
is suppressed by $\leq 0.2$ dex at all sampled luminosities.  The 
third curve corresponds to the wind-only model, whose LF is suppressed 
by a factor of $\approx2$ at all luminosities.  The effect is roughly 
as strong on bright sources partly because hierarchical growth spreads 
the impact on smaller objects into larger ones, and partly because 
we assume that outflows are active at all resolved masses.  Finally, 
including both outflows and an EUVB suppresses the galaxy abundance 
at all luminosities by a total factor of 2--3.

We also compare these predictions to recent constraints on the 
reionization-epoch LF~\citep{mcl10,bou11b}.  At $z=7$, the simulation 
including only an EUVB overproduces the observed LF in the regime
where the simulated and observed ranges overlap and the simulation's
Poisson errors are not large ($M_{1600}\approx-18$).  This confirms 
previous indications that observations require strong hydrodynamic 
feedback even at early times~\citep{dav06}.  Including outflows 
brings the predicted LF into agreement with observations, but now 
the EUVB's imprint is weak compared to current observational errors.  
The observed LF at $z=7$ has sufficiently large uncertainties that 
the wind-free simulations are only inconsistent with it at the 
1--3$\sigma$ level at any luminosity, but the fact that the offset 
is systematic argues strongly that it is real.  At $z=6$, the 
qualitative impact of different feedback effects is the same, but 
now observations at the faintest luminosities strongly favor models 
that include outflows.  As before, the imprint of a realistic EUVB 
at observable luminosities is weak compared to uncertainties.

\begin{figure}
\epsscale{1.1}
\plotone{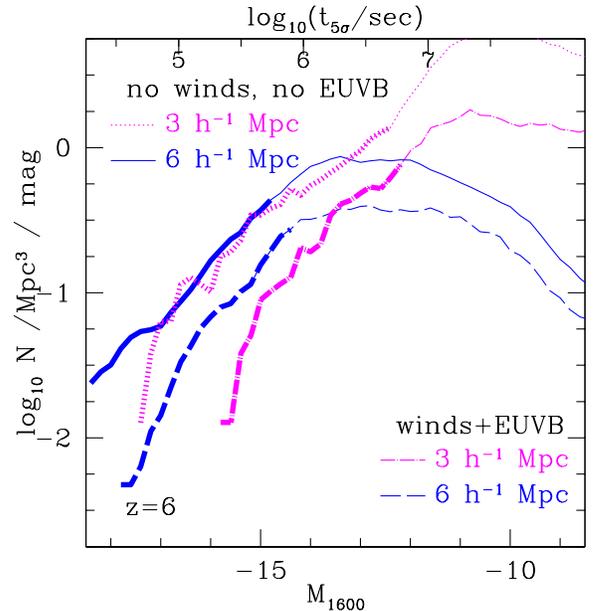}
\caption[]{The predicted LF at $z=6$.  Upper solid and dotted curves 
are from our fiducial and high-resolution volumes, respectively, in
simulations without any feedback. Lower long-dashed and dot-dashed
curves are from simulations of the same volumes that include both 
outflows and an EUVB.  For each simulation, heavy and light curves
indicate the predicted LFs brighter and fainter than the luminosity 
corresponding to our 64 star particle resolution limit, respectively.  
All curves
are smoothed with a boxcar 1 magnitude wide.  The top axis indicates 
the integration time for a $5\sigma$ detection in the F200W band of 
JWST in units $\log_{10}(t/\mathrm{sec})$.  Resolution limitations 
boost the luminosities in the fiducial simulation including feedback 
(lower long-dashed blue) by a factor of 2--3.  The predicted LF 
rises until at least $M_{1600}=-13$, a factor of 100 fainter than 
current observational limits at $z=6$ ($M_{1600}\approx-18$). 
}
\label{fig:lfUVr3r6}
\end{figure}

The predicted LFs in Figure~\ref{fig:lfUV} are affected by the resolution 
effects discussed previously.  In order to evaluate the strength of these
effects, we compare in Figure~\ref{fig:lfUVr3r6} the predicted LFs from 
our fiducial and high-resolution simulations at $z=6$.  Light and heavy 
curves show the LFs obtained from the complete simulated galaxy 
catalogs and when we exclude simulated galaxies with fewer than 64 star 
particles, respectively; we have previously argued that this corresponds 
to the minimum mass for reasonably converged star formation histories~\citep{fin06}.

Comparing the lower blue long-dashed and magenta dot-dashed curves suggests 
that the high-resolution winds+EUVB simulation yields galaxies that are 
roughly 1 magnitude fainter at constant number density than the low-resolution 
simulation with the same physical treatments.  In other words, halos at 
$z=6$ have 2--3 times less star formation at eight times higher mass 
resolution.  This numerical effect occurs in the presence of 
momentum-driven winds because higher-resolution simulations resolve star 
formation in lower-mass halos.  Lower-mass halos have stronger outflows, 
hence the massive halos into which they merge grow systematically 
baryon-deficient (Figure~\ref{fig:fbarmhalo}).  A qualitatively similar 
effect would occur even in the absence of 
feedback~\citep[as suggested by \S 3.9 of][]{ker09}, but 
momentum-driven outflows amplify it significantly.  To demonstrate
this, we also show the LFs from simulations that omit feedback (upper 
curves).  In this case, the low- and high-resolution simulations agree 
in the range where they overlap.  We conclude that outflows remain 
required in order to close the gap between the simulated and observed 
LFs at $z=6$, although increasing our resolution could allow us to 
decrease the wind normalization $\sigma_0$ (\S~\ref{sec:gadget}).  

In summary, both outflows and an EUVB suppress the normalization of
the predicted LF without significantly changing its shape for
absolute magnitudes brighter than $M_{1600}\leq-15$.  Over this range 
in luminosity, the observed LF favors models that include outflows 
without constraining the EUVB.  Resolution limitations artificially 
boost luminosities in our fiducial volume when outflows are included 
because outflows preferentially evacuate baryons from low-mass 
systems.  Correcting for this numerical effect could enable us to 
decrease $\sigma_0$ somewhat, but it would not remove the need for 
outflows.

\subsection{Turnover of the Luminosity Function}
The LF must flatten below some faint luminosity owing to the impact of
photoheating on photosensitive halos and eventually to the HI
cooling limit (Figure~\ref{fig:virials}).  Observing this feature 
would be a major step toward constraining the ionizing emissivity that 
galaxies contribute to cosmological reionization because current ``photon 
counting" reionization calculations are sensitive to the unknown abundance
of faint sources~\citep{cha08,mun11,bou11b}.

Theoretically, the turnover luminosity depends on the reionization 
history and the nature and strength of star formation feedback~\citep{bar00}.  
Regions that reionize sooner or develop a hotter IGM have a higher 
turnover luminosity.~~\citet{kul11} considered these effects by combining 
a semi-analytic model for reionization with a Jeans mass calculation of 
star formation suppression.  They found that the mass below which star 
formation stops varies between $10^{8\mbox{--}10}\msun$ at $z=8$ depending 
on the bias, corresponding to an absolute magnitude between -12 and -17.  
While their results are suggestive, their model had difficulty reproducing
the evolving LF.  Moreover, it did not treat gas flows or the growth of 
ionized regions in three dimensions, it did not account for the full
temperature history of a halo's environment, and it assumed the timescales
over which halos form stars and respond to an EUVB~\citep{dij04a} rather 
than computing them self-consistently.  Our numerical simulations are 
designed to overcome these limitations.

Observationally, the turnover luminosity at $z=6$ is constrained to 
be fainter than $M_{\mathrm{UV}}=-18$~\citep{bou10b}.~~\citet{mun11} 
have used a semi-analytic model to map this to a minimum star-forming 
halo mass of $10^{9.4^{+0.3}_{-0.9}}\msun$.  This is over an order 
of magnitude more massive than the HI cooling limit 
(Figure~\ref{fig:virials}).  Assuming that some star formation 
continues down to the HI cooling limit, the LF should continue to 
rise to lower luminosities.

Figure~\ref{fig:lfUV} shows that the predicted LF does not turn over 
in any model for luminosities brighter than $M_{1600}=-16$.  This 
constitutes a robust prediction that the observed LF will continue to 
climb to at least this luminosity at $z=6$~\citep[see also][]{jaa11}.  
In order to trace the 
predicted LF to even fainter limits, we refer to the high-resolution 
simulation including both outflows and an EUVB in Figure~\ref{fig:lfUVr3r6}.  
This simulation predicts that the LF continues its steep rise to at 
least $M_{1600}=-13$.  While the light curve flattens at fainter 
luminosities, the LF at luminosities fainter than -13 is dominated 
by galaxies with fewer than 64 star particles, hence the flattening 
likely owes largely to resolution limitations.  

Figures~\ref{fig:fbarmhalo}--\ref{fig:sfrmhalo} confirm
that an EUVB suppresses photosensitive halos, but 
Figure~\ref{fig:lfUVr3r6} indicates that even our high-resolution
simulation does not fully resolve star-forming galaxies at the 
turnover luminosity.  Nonetheless, we may bound the luminosity range 
within which a turnover is expected as follows: The turnover must
occur in the luminosity range corresponding to 
$\tvir=10^{4\mbox{--}5}$ K.  At $z=6$, Equation~\ref{eqn:sfrmh} and 
Table~\ref{tab:sfrmh} indicate that the extrapolated (that is, 
suppression-free) SFRs in halos with $\tvir=10^4$ K and $\tvir=10^5$ K
are $8.3\times10^{-4}$ and $0.10 \smyr$, respectively.  The predicted
relationship between $M_{1600}$ and SFR ($\smyr$) at $z=6$ is 
$M_{1600} = (-18.86\pm0.09) + (-2.56\pm0.06)\log(\mbox{SFR})$, hence 
the unsuppressed luminosities would be -11 and -16, respectively.  The 
bright end of this range may be overly conservative given that the LF
in our high-resolution simulation rises past -16 (Figure~\ref{fig:lfUVr3r6}).  
We therefore predict that the turnover at $z=6$ will occur between -11 
and -13.  Future work incorporating significantly larger dynamic range 
will be required in order to resolve the turnover numerically.

We may draw two conclusions from the faint ends in 
Figures~\ref{fig:lfUV}--\ref{fig:lfUVr3r6}.
First, hierarchical formation smooths the impact of feedback processes
on the LF dramatically such that neither outflows nor an EUVB creates
a characteristic feature above $M_{1600}=-13$.  Instead, hierarchical 
growth causes any feedback effect that suppresses low-mass halos to 
suppress the normalization of the entire LF.  We have directly verified 
that this is true at all $z<10$.  Second, we expect the LF to rise 
with decreasing luminosity until at least $M_{1600}=-13$ even in the 
presence of strong outflows and a realistic EUVB.  Detecting these faint 
objects at $z=6$ will require observations that probe a factor of 100 
deeper than the deepest observations to date, requiring 1--10 Msec on 
JWST (top axes).  

\section{Reionization Histories} \label{sec:reion}
The goal of the present work is to explore the relative impact of 
outflows and a self-consistently modeled EUVB on star formation in 
photosensitive halos.  In the process, we have simulated cosmological 
reionization.  Our simulations are not designed to resolve Lyman Limit 
systems or to capture the impact of density fluctuations on length 
scales that are large compared to our simulation volume, hence neither 
the ionization nor the recombination rates are expected to be 
numerically convergent.  Nevertheless, it is of interest to examine 
how the resulting reionization histories compare to observational 
constraints.  In this Section, we compare the volume-averaged neutral 
hydrogen fraction and EUVB amplitude to observational constraints from 
the Lyman-$\alpha$ forest and the cosmic microwave background.

\subsection{Integrated Optical Depth to Electron Scattering} \label{ssec:tau}
The probability that photons from the cosmic microwave background 
scatter off of free electrons following recombination is quantified 
by the integrated optical depth to Thomson scattering, $\tes$, which 
is a key observable that reionization models must confront.  
Reconciling the observed cosmic star formation history with $\tes$ 
is challenging~\citep{rob10,kul11,haa11} and depends on a number of 
assumptions including the evolving abundance of faint galaxies and 
the IGM recombination rate.  It is of interest to ask whether our 
simulations reproduce $\tes$ because they model these effects 
self-consistently.

\begin{figure}
\epsscale{1.0}
\plotone{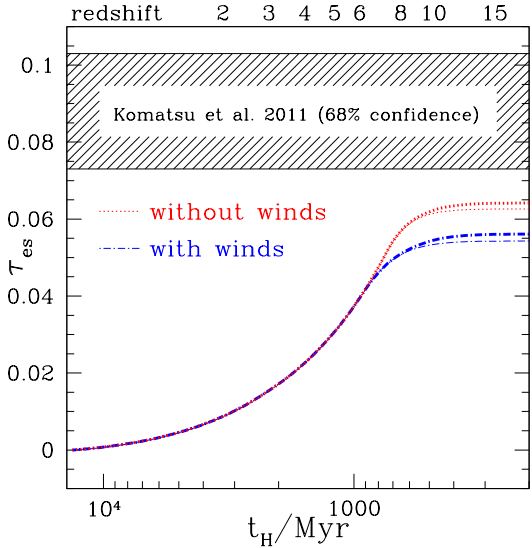}
\caption[]{The integrated optical depth to electron scattering $\tes$
as a function of the age of the universe.  Dotted red and dot-dashed 
blue curves indicate radiative transfer simulations without and with 
outflows, respectively.  Heavy and light curves 
compare results from simulations with radiative transfer grids with
cell widths of resolution 375 and 187.5 $\hkpc$, respectively.  
All curves include the estimated effect of Helium ionization.
Regardless of spatial resolution, helium, and the presence of 
outflows, simulations underproduce the observed optical depth 
of~\citet{kom11}.
}
\label{fig:tau}
\end{figure}

In Figure~\ref{fig:tau}, we show how outflows impact the integrated 
optical depth to electron scattering $\tes$ as a function of redshift.  
All curves assume that helium is singly ionized with the same ionization
fraction as hydrogen down to $z=3$, after which it is doubly ionized.
Upper red and lower blue curves indicate that simulations without (with)
outflows yield $\tes\approx0.063 (0.054)$, which lies below the 
observationally-determined 68\% confidence range of 
0.073--0.103~\citep{kom11}.  Doubling the spatial resolution of our 
radiation transport solver (light curves) suppresses $\tes$ slightly 
because simulations with higher resolution treat the small self-shielded 
regions (such as Lyman-limit systems) that dominate the IGM opacity at 
late times more accurately~\citep{fur05}, which slightly boosts the 
neutral hydrogen fraction (Figure~\ref{fig:xHIJ}).  The difference is 
small compared to the discrepancy with observations, hence spatial 
resolution limitations do not dominate the apparent underabundance 
of free electrons at early times.

Figure~\ref{fig:tau} may be compared with our post-processing 
calculations~\citep{fin09b}, where we also found values of $\tes$
in the range 0.05--0.07 depending on the choice of ionizing escape
fraction.  That work yielded higher values for both $\tes$ and
the EUVB amplitude while assuming a lower $\fesc$ because it did 
not account for subgrid recombinations.  Our new calculations 
treat the IGM opacity more correctly, which necessitates a higher 
$\fesc$ and yields a lower $\tes$ and EUVB amplitude.

\citet{pet10} used a different implementation of radiation hydrodynamics
into {\sc Gadget} to simulate reionization using $2\times256^3$ particles
within a $10\hmpc$ volume.  They accounted for galactic outflows using the 
model of~\citet{spr03}, hence their simulation is similar to ours, and they 
found $\tes=0.049$ (ignoring helium ionization).  When we recompute the 
$\tes$ from our r6wWwRT32 model omitting helium, we also obtain 
$\tes=0.0493$.  The remarkable agreement is probably coincidental given 
that our simulation considered a smaller volume with higher mass 
resolution, lower latent heat, a different outflow treatment, and higher 
$\fesc$.  Nevertheless, it remains intriguing that both simulations 
underproduce $\tes$.  \citet{pet10} attributed their underestimate 
partly to the missing contribution of ionized helium and partly to 
the fact that small cosmological volumes delay the onset of 
reionization~\citep{bar04}.  We estimate that helium contributes 
$\Delta \tes=0.005$, and our limited volume suppresses $\tes$ by 
$\approx0.003$ (assuming a redshift delay of 0.035 from Figure 3 
of~\citealt{bar04}).  Summing these 
effects, we estimate that accounting for both helium and our limited 
volume would boost the $\tes$ in our fiducial volume to 0.058--0.067, 
still shy of the observed range.  We will discuss other possible 
explanations for our low $\tes$ below.

\subsection{The Neutral Hydrogen Fraction and the Ionizing Background} \label{ssec:xHIJUV}

\begin{figure}
\epsscale{1.0}
\plotone{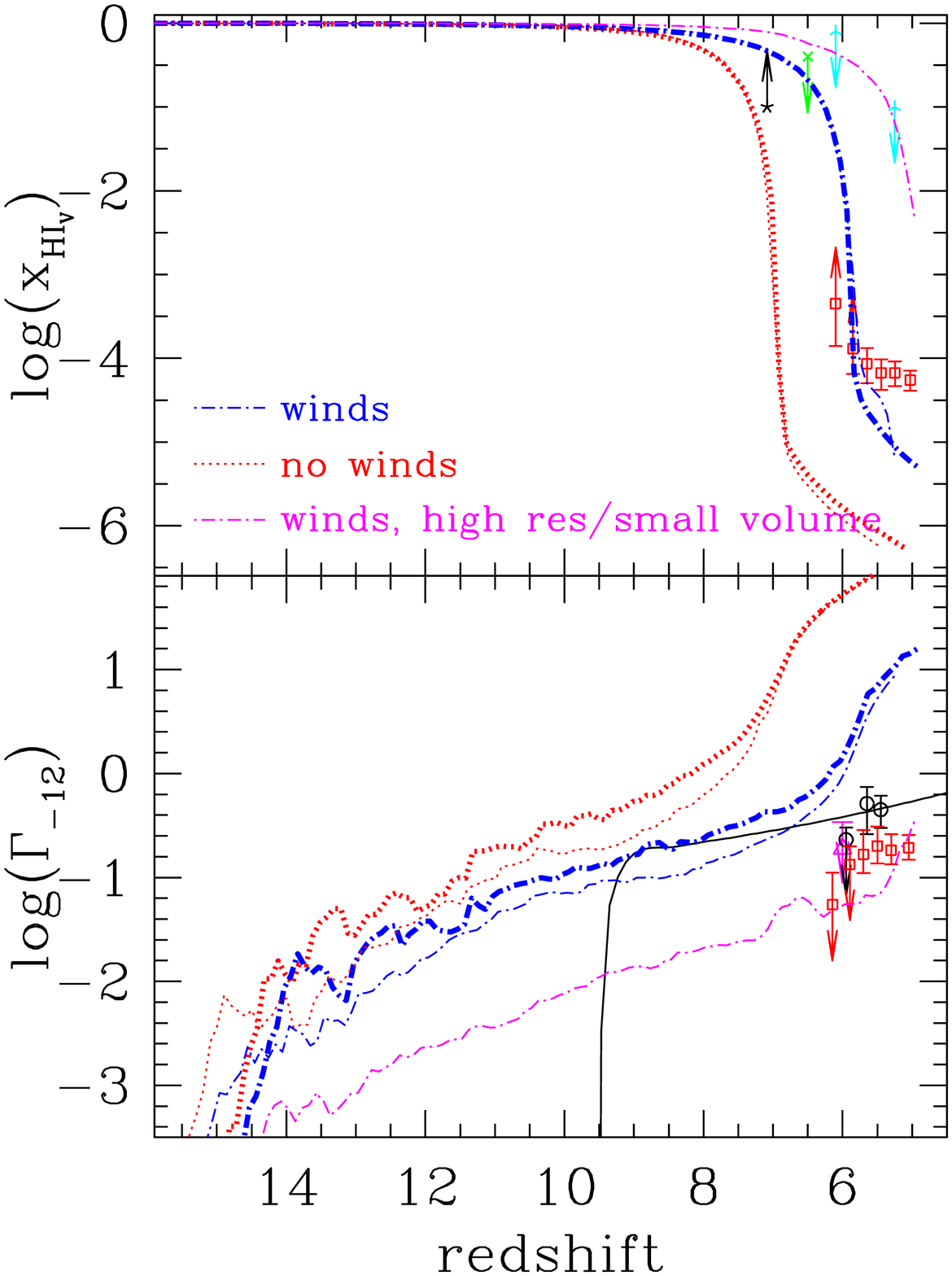}
\caption[]{\emph{(top)} Volume-averaged neutral hydrogen fraction
as a function of redshift. \emph{(bottom)} Ionization rate per hydrogen 
atom as a function of redshift.  Red squares are from~\citet{fan06},
the magenta triangle is from~\citet{bol07}, black circles are 
from~\citet{srb10}, and the (green, cyan, black) limits with 
(3, 4, 5)-pointed stars are from~\citet{kas11},~\citet{mcg11}, 
and~\citet{bol11} respectively.  We have updated the~\citet{mcg11} 
constraint at $z=5.25$ to 0.1 (private communication). The 
simulated curves' colors and styles are as in Figure~\ref{fig:tau}.  
The solid black curve indicates the observed 
galaxies+QSOs background HM01.  Simulations without outflows
readily reionize the Universe by $z=7$ while dramatically overproducing 
the observed EUVB.  Simulations including outflows complete reionization 
around $z=6$ while improving the agreement with the inferred strength
of the EUVB.  Our high-resolution/small-volume simulation completes
reionization at $z=5$ because it lacks bright sources and resolves 
more small-scale gas clumping.
}
\label{fig:xHIJ}
\end{figure}

We show in Figure~\ref{fig:xHIJ} how the volume-averaged neutral hydrogen
fraction and the ionization rate per hydrogen atom vary with redshift
and resolution.  If outflows are neglected (heavy red dotted), then the 
volume-averaged neutral hydrogen fraction drops below 0.01 around $z=7$ 
while the amplitude of the EUVB at $z=6$ is overproduced by a factor of 
100--1000.  Including outflows (heavy blue dot-dashed) delays the 
completion of reionization to $z=6$ while bringing both the neutral 
hydrogen fraction and the EUVB into improved (though imperfect) 
agreement with observations.  Our high-resolution, small-volume 
simulation (top magenta dot-dashed) does not complete reionization 
until $z\sim5$ owing to its high gas clumping and the lack of 
long-wavelength fluctuations.  
This history is at odds with the results of~\citet[][green arrow]{kas11}. 
These authors found that the Lyman-$\alpha$ LF of LAEs evolves more 
strongly during $z=6.5\rightarrow5.7$ than the UV continuum LF of LAEs.  
By assuming that reionization was complete at $z=5.7$, they used
the evolving Lyman-$\alpha$ equivalent width to derive a
volume-averaged neutral fraction $\xhiv$ at $z=6.5$ of $\xhiv\leq0.4$.  
This suggests that our small volume reionizes too late, although the 
constraint suffers from uncertainties associated with cosmic variance 
and the unknown intrinsic Lyman-$\alpha$ LF of LAEs.  Our small volume 
also conflicts with the commonly-accepted reionization redshift 
$z=6$~\citep{fan06}.  Note, however, that constraints from the 
Lyman-$\alpha$ forest generally assume a spatially-homogeneous EUVB 
(although see~\citealt{srb10}).  This assumption could lead to biased 
inferences at early times.~~\citet{mcg11} sidestepped the problem by 
obtaining an upper limit to $\xhiv$ from the dark pixel fraction in 
the Lyman-$\alpha$ forest (cyan arrows).  This maximally-conservative 
measurement yields much higher limits that are in fact consistent 
with our small volume.  However, given that many dark pixels likely 
correspond to regions that are largely ionized, it remains probable 
that our small volume requires a higher $\fesc$ in order to yield a 
realistic reionization history.

Our overall reionization histories could be sensitive to the spatial
resolution of our radiation transport solver.  To explore this, we 
use heavy and light curves to show reionization histories from
simulations that discretize the EUVB using $32^3$ and $16^3$ grid 
cells, respectively.  Doubling the spatial resolution of our radiation
transport solver (light curves) suppresses the EUVB amplitude by
a factor of $\sim$a few at early times without appreciably changing
the volume-averaged neutral fraction.  We expect its impact on our 
results to be modest.  

It is important to note that, by itself, Figure~\ref{fig:xHIJ} yields only
a joint constraint on the outflow amplitude $\sigma_0$ and the ionizing 
escape fraction $\fesc$ because increasing one can be compensated by 
decreasing the other.  Breaking this degeneracy requires galaxy 
observations, which constrain the outflow amplitude $\sigma_0$.  We have 
previously shown that $\sigma_0=150\kms$ suppresses the predicted galaxy 
LF into agreement with observations~\citep{dav06}.  Figure~\ref{fig:xHIJ} 
shows that combining this assumption with $\fesc=50\%$ yields a model 
that simultaneously reproduces observations of galaxies at $z=$6--8 and 
the commonly-accepted overlap redshift $z=6$~\citep{fan06}.  These 
considerations illustrate the crucial role of galaxy observations in 
constraining reionization models.

\subsection{Discussion} \label{ssec:params}

Figure~\ref{fig:tau} suggests that the ionized volume fraction is too
low at early times while Figure~\ref{fig:xHIJ} indicates that the 
ionization rate is too high at late times.  At the same time, we have
shown that our current wind model yields reasonable agreement with the 
observed UV continuum LF of galaxies at $z=6$ 
(\citealt{dav06};~\citealt{fin11}; Figure~\ref{fig:lfUV}), which is 
proportional to the ionizing emissivity.  As a guide for future work 
toward the goal of 
devising a self-consistent model for cosmological reionization and 
feedback effects, it is instructive to consider how our parameter 
choices could impact our results.  The free physical parameters are 
the latent heat per photoionization 
$\epsilon_{\mathrm{HI}}$, the ionizing escape fraction $\fesc$, and the 
outflow amplitude $\sigma_0$.  The free numerical parameters are the 
simulation volume and the resolution of our SPH and radiation transport 
solvers.  We now discuss each of these in turn.

Increasing the latent heat per photoionization $\epsilon_{\mathrm{HI}}$ 
heats the IGM and boosts Jeans smoothing.  This simultaneously 
suppresses the IGM recombination rate (and through it the neutral 
hydrogen fraction) while only modestly suppressing the predicted 
galaxy LF (and through it the amplitude of the EUVB).  For 
example,~\citet{pet10} show that increasing $\epsilon_{\mathrm{HI}}$ 
from $6.4\rightarrow30$ eV accelerates reionization by 
$\Delta z \approx 0.5$ while roughly doubling the EUVB amplitude.
Our current choice $\epsilon_{\mathrm{HI}} = 0.3$ Ryd is motivated 
directly by the metallicity-dependent stellar continua that our 
simulations predict.  The good agreement that similar simulations 
have achieved with the observed history of high-redshift star 
formation~\citep{dav06,fin11} and the metal abundance in 
galaxies~\citep{fin08,dav11b} and in the IGM~\citep{opp06} argues 
that our predicted stellar continua are realistic.  Increasing 
$\epsilon_{\mathrm{HI}}$ would therefore run against insight from 
existing observations.  Moreover, it could lead to implausible IGM 
temperatures.  For example,~\citet{pet10} find that assuming a latent 
heat per photoionization of 30 eV (compared to our 4.08 eV) yields 
photoionized regions with temperatures in excess of $10^6$K, which may 
be difficult to reconcile with the tight observed association between 
cool gas and Lyman-Break galaxies~\citep{wei09,ste10}.  It would
also violate agreement with the observed distribution of CIV line
widths, which does not show evidence for such significant 
heating~\citep{opp06}.  Nevertheless, a modest increase in 
$\epsilon_{\mathrm{HI}}$ could improve agreement without violating 
existing constraints.  Such an increase could be associated with 
spectral filtering through interstellar media. 

Increasing $\fesc$ causes reionization to occur sooner by strengthening
the EUVB~\citep{fin09b}.  This means that increasing $\fesc$ would boost 
$\tes$ into better agreement with observations of the cosmic microwave 
background (Figure~\ref{fig:tau}), but at the cost of exacerbating the 
disagreement with IGM observations at $z\sim6$ (Figure~\ref{fig:xHIJ}).  
Decreasing $\fesc$ in the case of the no-wind model would delay 
reionization, weaken the EUVB, and increase the neutral fraction.  
This would improve its agreement with observations in 
Figure~\ref{fig:xHIJ}, but at the expense of further suppressing $\tes$.  
The only way in which $\fesc$ could simultaneously improve the
discrepancies in Figures~\ref{fig:tau}--\ref{fig:xHIJ} would involve
a mass dependence such that $\fesc$ is larger in low-mass systems whose
star formation halts by $z=6$~\citep[similar to][]{ili07}.

Weakening outflows (by decreasing $\sigma_0$) accelerates reionization
in two ways.  First, it increases the SFR.  As can be 
seen by examining our no-wind simulation---which corresponds to the 
extreme case $\sigma_0 = 0$---this boosts the ionizing emissivity.  A 
weaker effect comes from the fact that weakening outflows at constant 
$\fesc$ decreases the amount of absorbing gas in the IGM.  This can 
be seen by comparing the number of ionizing photons per hydrogen at 
overlap, the last column in Table~\ref{table:sims}: No-wind simulations 
require $\approx4$ photons per hydrogen to reionize whereas wind models 
require $\approx5$.  Intuitively, outflows relocate absorbing gas from 
the ISM into the halo so that removing outflows at constant $\fesc$ is 
equivalent to boosting $\fesc$.  Both effects boost the EUVB amplitude
and $\tes$.  Unfortunately, the EUVB is already too strong, and 
weakening outflows also causes simulations to overproduce the observed 
LF of galaxies~\citep{dav06}, which is now
well-constrained.  In fact, our most recent simulations suggest that 
this parameter is more likely too low (too little suppression) than too 
high~\citep{fin11}.  Hence it is not likely that boosting $\tes$ by 
decreasing $\sigma_0$ would lead to overall improved agreement.

Increasing the simulation volume at constant mass resolution causes 
reionization to begin sooner by more completely sampling the rare
high-$\sigma$ peaks that were the first to collapse~\citep{bar04}.  
This boosts $\tes$ without affecting the galaxy LF or the IGM 
temperature at late times.  We estimate that this could increase 
$\tes$ by 0.003--0.004, which helps but may not be enough to bring 
predictions and observations into agreement.

Increasing the mass resolution would help in two ways.  First, it 
would boost the IGM recombination rate by resolving small self-shielded
systems, limiting the mean free path to ionizing photons at late 
times~\citep{ili05,aub10}.  The potential 
for improvement can be seen by comparing the light blue and magenta 
dot-dashed curves in the bottom panel of Figure~\ref{fig:xHIJ}.  These 
two simulations adopt the same physical treatments, but the 
higher-resolution simulation's (magenta) increased mass and spatial 
resolution significantly decrease the mean free path to ionizing photons
while its small volume lacks the sources that dominate star formation 
following $z=13$ (Figure~\ref{fig:madau}).  The net result is that
the EUVB amplitude is lower even before its SFRD 
is underproduced while the completion of overlap is delayed until $z=5$.

The second benefit of increased mass resolution would result from higher 
ionizing emissivity at early times.  To see this, note that 
Figures~\ref{fig:tau}--\ref{fig:xHIJ} imply a need for extra ionizations 
at early times but not later on.  This could be achieved through a 
population of low-mass halos that are active at early times but 
suppressed by a mature EUVB~\citep{ili07}.  Our fiducial simulations 
account completely for star formation in halos above the HI cooling 
limit at $z=7$ of $2\times10^8\msun$, but they fail to account for 
the star formation that must have occurred in lower-mass halos at 
earlier times, when the HI cooling limit was lower 
(Figure~\ref{fig:virials}).  Increasing our mass resolution would 
boost star formation prior to $z=7$ while leaving it unchanged at 
later times, thus increasing $\tes$ without affecting the EUVB at 
$z=6$.  Unfortunately, for the reasons mentioned previously,
increasing our mass resolution would simultaneously increase the 
ionizing emissivity and the recombination rate, hence the net 
effect on $\tes$ requires numerical modeling.

We may estimate how much increasing our dynamic range would increase 
$\tes$ by taking the larger of the electron abundances predicted by 
our $3\hmpc$ and $6\hmpc$ simulations at each redshift:
\begin{eqnarray}
\Delta \tes = c \sigma_T \int (\mathrm{max}(n_{e,\mathrm{lr}},n_{e,\mathrm{hr}}) - n_{e,\mathrm{lr}}) dt
\end{eqnarray}
Here, $n_{e,\mathrm{lr}}$ and $n_{e,\mathrm{hr}}$ represent the
electron abundances in the large, low-resolution and small, high-resolution
simulations, respectively.  Carrying out the integral, we find that
$\Delta \tes = 7\times10^{-6}$, with all of the excess scattering 
occurring during the redshift range $z=$15--22 since the smaller 
volume has a lower SFRD at later times 
(Figure~\ref{fig:madau}).  This correction is small compared to the 
discrepancy with~\citet{kom11}.  In detail, it is underestimated owing
to the decreased presence of long-wavelength density fluctuations in
our smaller comparison volume.  While a full accounting requires 
simulations with increased dynamic range, we conclude for the present 
that increased mass resolution would significantly improve the predicted 
EUVB amplitude and IGM ionization fraction without completely correcting 
the low $\tes$.

The final way in which we could reconcile our models with observations
involves invoking an additional population of ionizing sources that
is active at early times such as Population III stars or 
miniquasars~\citep{mad04}.  In order for miniquasars to bring $\tes$ 
into agreement with observations, they must provide an extra 
$\Delta \tes = 0.019$.  This could be achieved through a constant 
ionized fraction of 0.15 from $z=20\rightarrow10$.  Maintaining this 
ionized fraction against recombinations would require $2.4\times10^{-19}$ 
ionizations cm$^{-3}$s$^{-1}$ (proper units) for a clumping factor 
of 10, which is a characteristic value from our simulations at an
ionized volume fraction of 10\%.  The ionizing luminosity of 
Eddington-limited accretion by a black hole of mass $M_{BH}$ is 
$\sim10^{51} (\fesc/1) (M_{BH}/1000 M_\odot) $ s$^{-1}$~\citep{mad04,dij06},
hence this level of ionization could be supported by a population
of $1000\msun$ black holes with a comoving space density of 
$\sim$1--2 Mpc$^{-3}$.  This corresponds roughly to the space 
density of dark matter halos more massive than $5\times10^8\msun$ at 
$z=6$, hence such a population could be hosted entirely by 
atomically-cooled halos.

In summary, Figures~\ref{fig:tau}--\ref{fig:xHIJ} indicate that our new
simulations underproduce the ionization rate at early times while 
overproducing it at late times.  Increasing $\epsilon_{\mathrm{HI}}$, 
the comoving volume, or the mass resolution would yield simultaneous 
improvement in both respects.  However, self-consistency argues against 
significantly higher values of $\epsilon_{\mathrm{HI}}$; simple estimates 
indicate that the impact of our limited volume is small; and a 
high-resolution simulation suggests that increasing the mass resolution 
boosts the IGM recombination rate enough to counter the boosted 
ionizing emissivity at early times.  Meanwhile, increasing $\fesc$ or 
decreasing $\sigma_0$ would improve $\tes$ while degrading agreement 
with observations of galaxies and the IGM at $z=6$.  We conclude that
simultaneously matching observations of galaxies, the IGM, and the 
cosmic microwave background may require us to modify assumptions 
regarding feedback, the nature of the ionizing sources, or the ionizing 
escape fraction.  We plan to consider these possibilities in future 
work.  The important point for our present purposes is that our 
simulations yield sufficiently realistic reionization histories that 
we may use them to gain qualitative insight into how outflows and 
photoheating modulated galaxy growth during the reionization epoch.

\section{Summary} \label{sec:summary}

We have used a suite of cosmological radiation hydrodynamic simulations 
to study the impact of galactic outflows and photoheating
by a self-consistent, spatially-inhomogeneous EUVB on star formation 
during the reionization epoch.  The major improvements of our work over
previous radiation hydrodynamic studies~\citep[for example,][]{gne00,pet10}
are that our model includes a treatment for galactic outflows that has
previously been tested extensively against observations of galaxies 
and the IGM from $z=0\rightarrow7$ (\S~\ref{sec:intro}), and that it
essentially resolves the HI cooling limit at all relevant redshifts.  
It is the first study to demonstrate agreement with the observed UV 
continuum LF, achieve reionization by $z=6$, and resolve the majority 
of the star formation that occurs in photosensitive halos.  Our major 
results are as follows:

\begin{itemize}
\item By $z=6$, a self-consistent, inhomogeneous EUVB significantly
suppresses the baryon reservoirs and SFRs in halos less massive than 
$3\times10^9\msun$.  Meanwhile, the star formation activity in halos 
more massive than $3\times10^9\msun$, which host currently observable 
galaxies,
is essentially unaffected.  This suggests that an EUVB acting alone 
cannot suppress the UV LF into agreement with observations at $z=6$.
\item Momentum-driven outflows can suppress the EUVB and the LF into 
improved agreement with observations without preventing Population 
I--II star formation from reionizing the Universe by $z=6$ as long 
as we assume that, on average, 50\% of ionizing photons escape from 
the low-mass galaxies that drive reionization.
\item Even in the presence of strong outflows, an inhomogeneous EUVB
allows photosensitive halos to contribute up to $\sim50\%$ of all 
ionizing photons throughout the reionization epoch.
\item For halos in the mass range $M_h=10^{8.2\mbox{--}10.2}\msun$,
SFR scales as $M_h^{1.3\mbox{--}1.4}$.  If the escape fraction is
constant, this implies substantially more power in 21-centimeter 
fluctuations on large scales (0.1 h Mpc$^{-1}$) than would be 
expected if $\mbox{SFR} \propto M_h^{1.0}$.
\item High-resolution simulations indicate that the LF continues 
to rise steeply until at least $M_{1600}=-13$.  This implies that 
reionization was dominated by an abundant population of faint 
galaxies that has not yet been observed.
\item Outflows and an EUVB couple nonlinearly in two ways: First, 
outflows weaken the EUVB, boosting star formation in photosensitive 
halos and leading to overall de-amplification of suppression at 
early times.  Second, they amplify the impact of a given EUVB on 
photoresistant halos as suggested by~\citet{paw09}, leading to overall 
amplified suppression of the cosmic comoving SFRD near the end 
of reionization.  By $z=6$, overall suppression of star formation is 20\% 
greater than would be expected from the two effects acting separately.
\item Simulations that achieve reionization at $z=$6--7 (that is, our
fiducial volume) generically underpredict the optical depth to Thomson 
scattering while slightly overproducing the inferred amplitude of the 
EUVB and the volume-averaged ionized hydrogen fraction 
at $z=6$.
\item Discrepancies between observations and models could be alleviated
through adjustments to physical or numerical parameters.  In general,
however, our physical parameters are either indirectly constrained 
observationally or could only improve agreement significantly with one 
observable at the expense of another.  Simple estimates suggest that 
improved dynamic range would not boost the predicted $\tes$ into 
agreement with observations although it would help.  
\item Observations thus suggest the 
need for an additional physical scaling that preferentially boosts the 
ionizing emissivity at early times such as a mass-dependent ionizing 
escape fraction.  Alternatively, an additional population of ionizing 
sources such as miniquasars could have provided the extra ionizations.
\end{itemize}

While our results represent a significant step forward in the modeling 
of reionization-epoch star formation, our predictions are not yet 
completely converged with respect to mass, spatial, or spectral 
resolution.  The need for higher mass resolution can be seen in our 
predicted UV luminosity function, which is slightly high owing to 
the delayed onset of star formation at finite mass resolution 
(Figure~\ref{fig:lfUVr3r6}).  We have also argued that improving our 
mass resolution would alleviate tensions with inferences from the 
cosmic microwave background (\S\ref{ssec:params}).

The need for improved spatial resolution within our radiation transport 
solver can likewise be seen in two ways.  First, comparing 
the number of photons required to achieve reionization in our r6wWwRT16 
and r6wWwRT32 simulations (Table~\ref{table:sims}, column 6) reveals 
that the higher-resolution simulation ``consumes" 4\% fewer photons.  
This is because the low-resolution simulation smooths over regions that 
are in reality self-shielded, thereby overestimating gas clumping.
Second, the fact that the light and heavy blue dot-dashed curves in
Figures~\ref{fig:fbarmhalo}--\ref{fig:sfrmhalo} are not coincident in
the photosensitive range shows that our simulations do not completely
resolve the self-shielded regions that host photosensitive halos.  
A convergent prediction would require RT cell widths that are smaller 
than the length scales of Lyman Limit systems ($\sim10$ physical 
kpc;~\citealt{sch01}), which is four times smaller than the RT cells
in our r6wWwRT32 simulation.  Current computational resources already 
allow us to augment our dynamic range somewhat.  This will improve our 
estimate of the IGM recombination rate, confine the EUVB more completely 
to overdense regions at early times, and resolve the earliest stages of 
star formation in photosensitive halos.  

Another potentially important effect involves fluctuations in the 
ionizing background on still smaller scales.  For 
example, the interstellar radiation field within an individual 
galaxy can dominate the EUVB in wavelengths to which its ISM is
optically thin~\citep{gne10a,can10}, but our current simulations do 
not treat it separately from the EUVB.  Consequently, they may
overestimate the amount of feedback that is required in the form
of outflows in order to reproduce the observed LF.  This process
would also assist in alleviating lingering tensions between the 
predicted and observed gas fractions of low-mass galaxies, which
remain too low at low redshifts in models similar to this 
one~\citep{dav11a}.  We will consider in future work whether this
effect can be incorporated into our subresolution treatment for
star formation.

Upgrades to our radiation transport solver allowing multifrequency
calculations and the addition of a background from quasars will result
in a more accurately-modeled EUVB.  Accounting for the longer mean
free path of higher-energy photons would raise the IGM temperature in 
ionized as well as neutral regions~\citep{abe99,ili06b,tit07}, modifying the
relative roles of photosensitive and photoresistant halos.

Finally, recent observations suggest that the way in which dense 
gas collapses into stars could include dependencies such as the 
gas metallicity~\citep{gne10b,kru11} and the stellar mass 
density~\citep{shi11}.  Incorporating these dependencies could 
modify the predicted SFR-$M_h$ relation by delaying star formation 
activity in photosensitive halos, which would again boost the 
expected power spectrum of 21 centimeter fluctuations at large 
scales.

\acknowledgments
The authors thank Steven Furlanetto, Chael Kruip, Joey Mu{\~n}oz, 
Peng Oh, and Moire Prescott for helpful discussions.  We thank 
Volker Springel for making {\sc Gadget-2} public.  We thank John
Wise and Ilian Iliev for sharing their test results with us.  We
thank the referee for many thoughtful suggestions that improved 
the draft.  Our 
simulations were run on the University of Arizona's Xeon cluster.
Support for this work was provided by the NASA Astrophysics Theory
Program through grant NNG06GH98G, as well as through grant number
HST-AR-10647 from the SPACE TELESCOPE SCIENCE INSTITUTE, which is
operated by AURA, Inc. under NASA contract NAS5-26555.  Support for
this work, part of the Spitzer Space Telescope Theoretical Research
Program, was also provided by NASA through a contract issued by the
Jet Propulsion Laboratory, California Institute of Technology under
a contract with NASA.  This work was also supported by the National 
Science Foundation under grant numbers AST-0847667 and AST-0907998. 
Computing resources were obtained through grant number DMS-0619881 
from the National Science Foundation.  KF acknowledges support 
from NASA through Hubble Fellowship grant HF-51254.01 awarded by the 
Space Telescope Science Institute, which is operated by the Association
of Universities for Research in Astronomy, Inc., for NASA, under
contract NAS 5-26555.\linebreak

\appendix
\section{Test of Multiple Sources}\label{sec:tests}
Test 4 of~\citet{ili06b} involves evolving the ionization fronts from 16 
sources embedded in a static cosmological density field, hence it tests 
the code's ability to resolve the impact of density fluctuations on the shape 
of the ionization front.  It also tests the code's treatment for photoheating 
and radiative cooling.  For multifrequency techniques it tests 
spectral filtering, but as our code is currently monochromatic, it 
illustrates the systematics associated with neglecting the ability
of high-energy photons to preheat neutral regions~\citep{ili06b,tit07}.

We represent the density field using a uniform grid of SPH particles 
whose masses vary in such a way as to reproduce the test requirement in 
the absence of SPH smoothing.  During run-time, our code extracts the 
gridded densities from the SPH particles using a ``clouds-in-cells"
approach.  This smooths the density field on a length scale of $\sim3$ 
times the mean particle separation.  Consequently, the opacity field 
that our radiation transport solver ``sees" does not exactly reflect 
the test requirement, and we do not expect our results to agree 
quantitatively with those of other codes.  We have verified that 
increasing the number of SPH particles per grid cell from 1 to 8 
yields ionized regions whose shapes agree more closely with reference 
results.  We enclose the test
volume in 4 layers of opaque boundary cells that prevent photons from 
``wrapping around" because our code is optimized for periodic 
boundary conditions.  The latent heat per photoionization is 1.177 
Ryd, which is appropriate for a $T=10^5$ K blackbody in the 
optically-thick approximation.  We smooth the Eddington tensor with a 
27-cell tophat filter in order to suppress rapid spatial fluctuations 
in the radiation pressure tensor.  This step ensures photon 
conservation at the cost of rendering the radiation field more diffusive.  
We have found empirically that smoothing is not required in our current
cosmological simulations because their spatial resolution is generically 
much lower than that of Test 4, preventing the appearance of sharp peaks 
in the radiation pressure tensor.  However, future cosmological 
simulations at higher spatial resolution may require smoothing.
We run the test in an 
expanding Universe at $z=8.84922$ with $\Omega_b = \Omega_M = 1$, and
we adjust $H_0$ and the box size so that the proper density and 
volume match the required test conditions.  Strictly speaking, 
Test 4 is meant to be a test of static density fields and should not 
be run in an expanding frame.  However, its duration of 0.4 Myr is 
small compared to the Hubble time at this redshift, hence 
cosmological effects are negligible.

\begin{figure}
\epsscale{1.0}
\plottwo{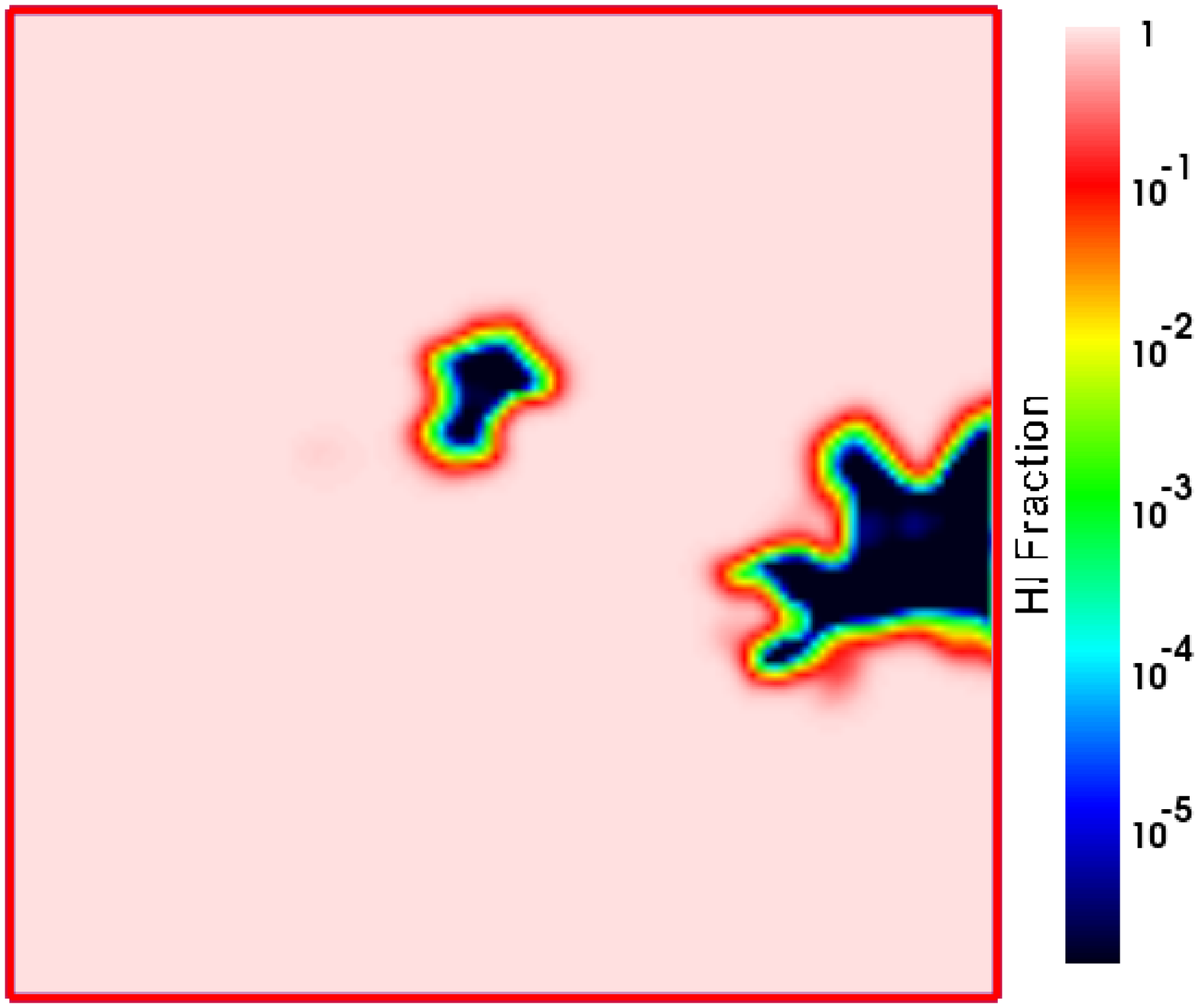}{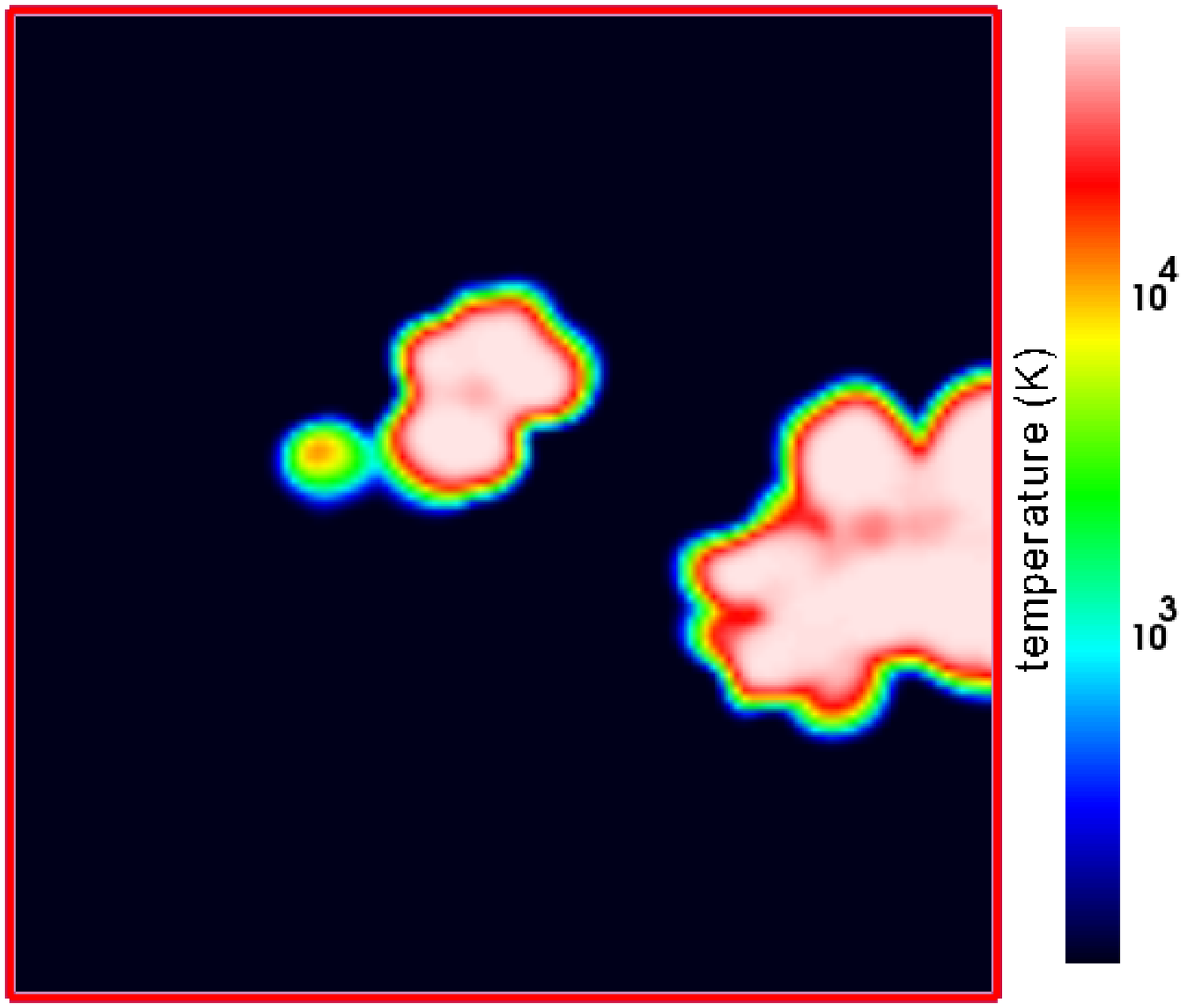}
\plottwo{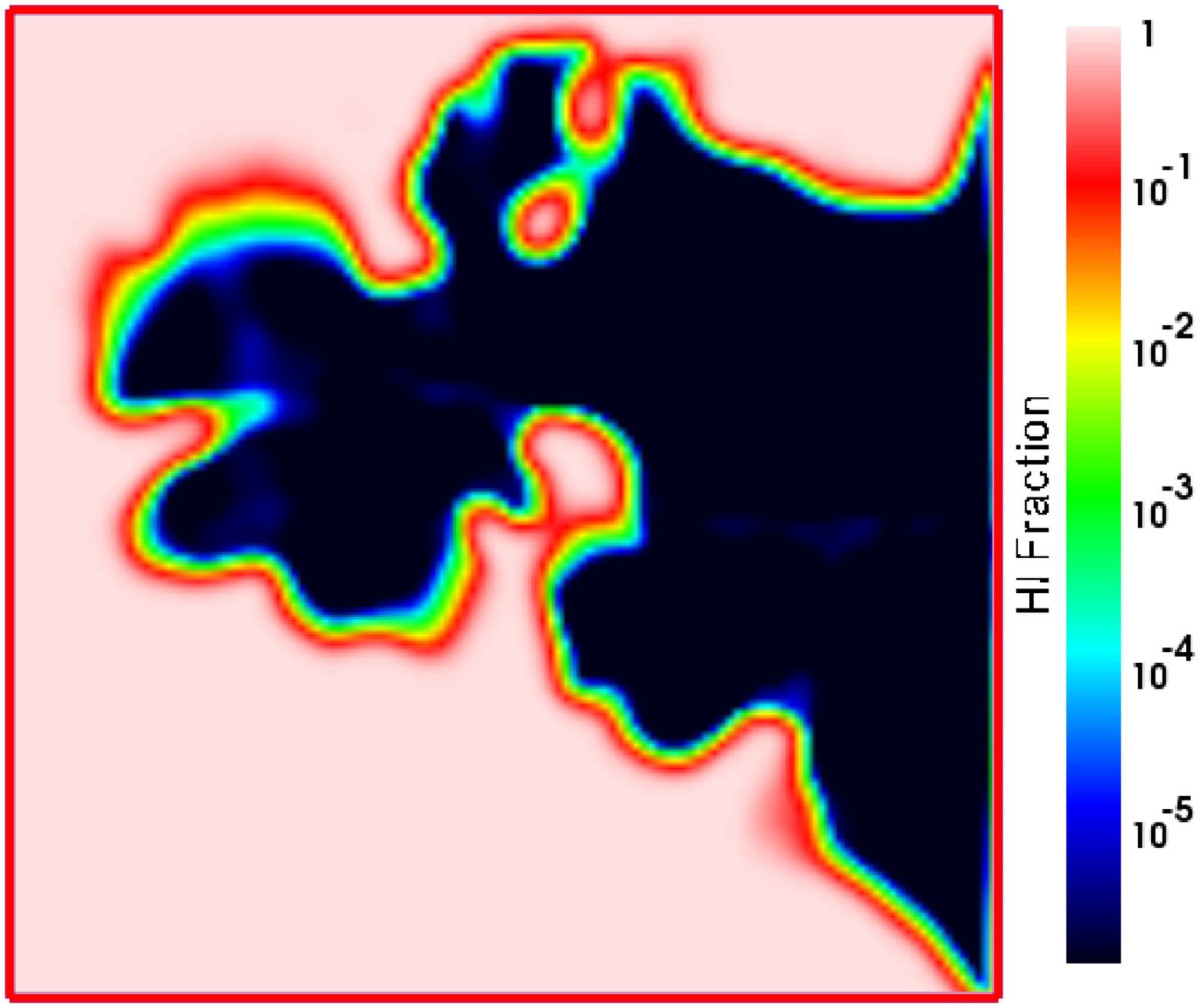}{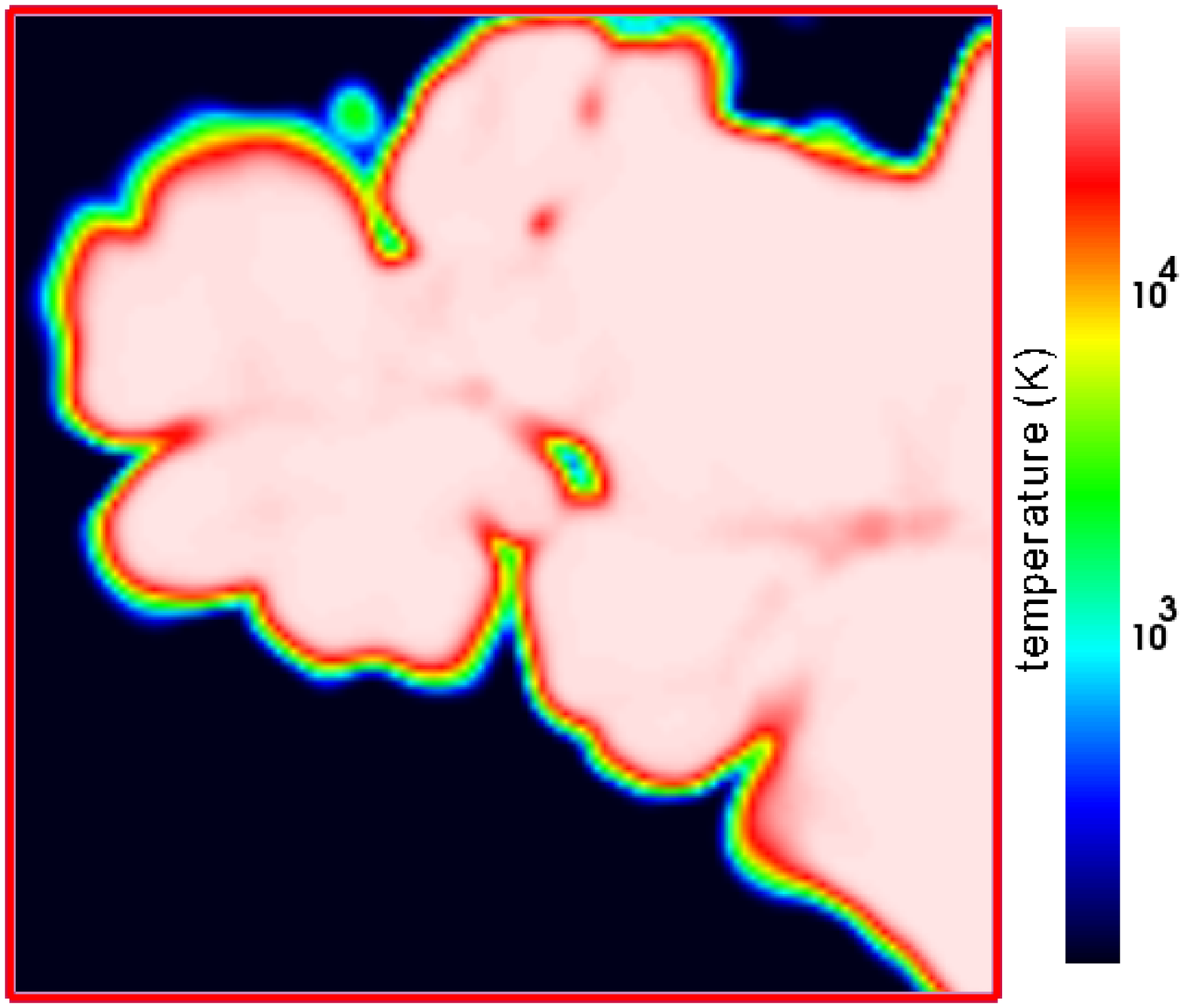}
\caption[]{Ionization state \emph{(left)} and temperature \emph{(right)} 
of the Test 4 case at $t = 0.05$ Myr \emph{(top)} and $t = 0.2$ Myr \emph{(bottom)}.  
This test was run with $256^3$ SPH particles and $128^3$ RT cells.
The ionization front is significantly smoother than those produced by 
other codes~\citep{ili06b,wis11,paw11}
owing to the diffusive nature of our radiation transport solver, and the 
IGM behind the ionization front remains cold because our monochromatic 
solver does not treat spectral hardening.  Nonetheless, the volume and 
temperature of the ionized region are reasonable, indicating that the 
code accurately evolves the IGM's thermal and ionization evolution.
}
\label{fig:test4_maps}
\end{figure}

We show maps of the neutral hydrogen fraction and temperature in
a slice of the simulation volume in Figure~\ref{fig:test4_maps}.
The ionization maps (left column) show similar morphologies to
the results from other codes (see, for example, Figure 19 
of~\citealt{wis11} or Figure 10 of~\citealt{paw11}).  This broad 
agreement indicates that our approach accounts reasonably well for 
fluctuations in the EUVB on length scales that are large compared 
to individual cells.  In detail, our approach yields smoother ionization 
fronts than ray-tracing codes partly because SPH smooths the density 
field on scales comparable to the mean particle separation, and partly 
because the moment method smooths the radiation field on scales 
comparable to the cell length.  While this motivates further work
in order to capture fluctuations at length scales that are closer
to the size of the RT grid cells, we do not expect it to alter 
the interaction between galaxies and the EUVB when averaged over
cosmological volumes.

\begin{figure}
\epsscale{0.5}
\plotone{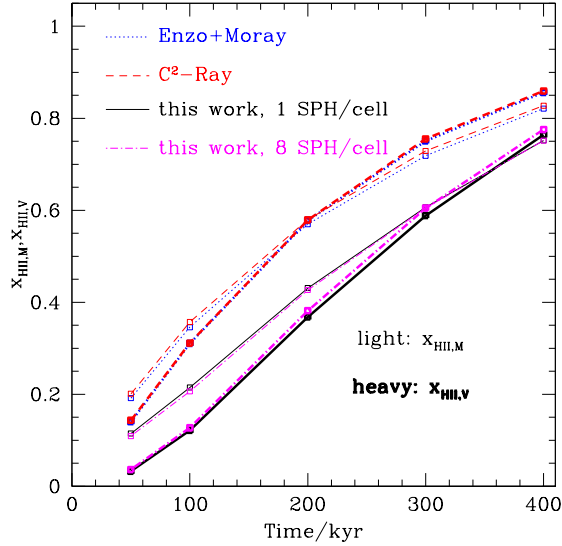}
\caption[]{Mass-weighted (light) and volume-weighted (heavy) 
ionized fraction versus time in our approach (solid black) and 
using {\sc c$^2$-ray} (dashed red) and {\sc enzo+moray} (dotted 
blue).  Solid black and dot-dashed magenta curves indicate test 
runs with 1 and 8 SPH particles per grid cell, respectively.  
The topology of reionization and the ionized fractions are both 
sensitive to the density field's accuracy.  Broadly, our 
predicted ionized fractions are somewhat lower than in the 
reference codes, indicating that our method predicts a slower 
reionization history in this test case.  
}
\label{fig:test4_fracs}
\end{figure}

For a more quantitative comparison, we show in 
Figure~\ref{fig:test4_fracs} how the volume-averaged and 
mass-averaged ionized fractions evolve in our own test (black 
solid) and in tests conducted using {\sc enzo+moray}~\citep{wis11} and 
{\sc c$^2$-ray}~\citep{mel06}.  Two kinds of differences are apparent.  
First, the predicted 
reionization topologies differ.  The ratio $\xhiim/\xhiiv$ exceeds 
unity in all codes at early times, indicating that the initial stages 
of reionization are robustly ``inside-out"~\citep{ili06a}.  The 
ray-tracing codes switch to an ``outside-in" topology by the time the 
ionized fraction reaches 60\%.  This indicates that ionization fronts 
have arrived in underdense regions and is reminiscent of the 
``inside-outside-middle" topology discussed in~\citet{fin09b}.  
Meanwhile, our moment approach maintains a purely inside-out
topology until the ionized fraction reaches nearly 70\%, indicating
significantly stronger ionization-front trapping in overdense 
regions.  Increasing the number of SPH particles per grid cell moves 
the transition to earlier times, suggesting that
the difference between the predicted topologies may owe largely to 
our smoothed density field.  The second difference involves the 
reionization rate.  
Our method predicts that reionization proceeds more slowly than in 
the reference codes, with the neutral fractions $\sim10\%$ higher 
at the end of the test.  Comparing the 
solid black and dot-dashed magenta curves reveals that increasing
the accuracy with which we capture the required density field
weakens ionization-front trapping and accelerates reionization.  
However, the difference is slight, suggesting that increasing our 
spatial resolution indefinitely would not necessarily bring our 
results into agreement with the reference runs.

\begin{figure}
\epsscale{0.5}
\plotone{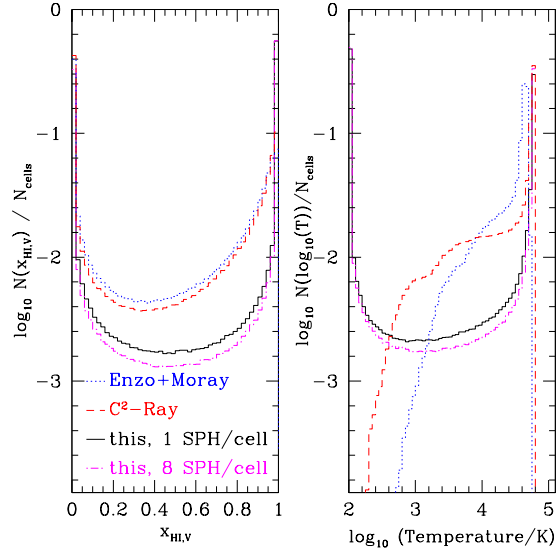}
\caption[]{Histograms of neutral hydrogen fraction (left) and
temperature (right) at 0.2 Myr.  Colors and line styles are as 
in Figure~\ref{fig:test4_fracs}.  Reionization occurs somewhat 
more slowly in our code as compared to the other codes, and the 
absence of high-energy photons leads to a systematically cooler 
IGM.
}
\label{fig:test4_probs}
\end{figure}

In order to explore our code's slower reionization rate further, we 
show histograms of neutral fraction and temperature at 0.2 Myr in 
Figure~\ref{fig:test4_probs}.  The distribution of neutral fractions 
has a similar shape in all four cases, but our code shows fewer cells 
in all bins of $\xhiv$ except the neutral bin $\xhiv=$0.98--1.0.  In 
fact, the fraction of cells in which $\xhiv=1$ is $\sim40\%$ in our 
test runs and $<10^{-3}$ in both reference runs.  This 
dominates the difference between our code and the reference 
codes in Figure~\ref{fig:test4_fracs}.   Accounting more completely 
for partial ionization by high-energy photons in neutral regions 
would improve agreement.  For similar reasons, our code does not 
pre-heat neutral regions, leading to an enhanced volume fraction at 
low temperature with respect to the reference codes.

In summary, our radiation hydrodynamic code yields reionization 
histories that are similar to results from other codes.  This implies 
that our approach accounts reasonably well for fluctuations in the EUVB 
on scales larger than an individual cell.  In detail, however, 
underdense regions are colder and more neutral in our code than in others.
This partly reflects the difficulty of translating a gridded density 
field into a grid of SPH particles.  Mostly, however, it simply reflects 
the absence of high-energy photons that broaden ionization fronts and 
preheat neutral regions.

\end{document}